\documentclass[a4paper,11pt]{article}
\usepackage{latexsym}	%Package for feynman diagrams.
\usepackage{amsmath}
\usepackage{amsthm}

\usepackage{ amssymb }
\def\ba{\begin{eqnarray}}
\def\ea{\end{eqnarray}}

\def\R{\hbox{\bf R}}

\def\ba{\begin{eqnarray}}
\def\ea{\end{eqnarray}}

\def\R{\hbox{\bf R}}
\def\lb{\label}
\def\be{\begin{equation}}
\def\ee{\end{equation}}

\theoremstyle{plain}
\newtheorem{theorem}{Theorem}%[section]
\newtheorem{prop}[theorem]{Proposition}%[section]
\def\R{\mathbb R}
\begin{document}
\baselineskip0.25in
\title{About the Cauchy problem in Stelle's quadratic gravity}
 \author{Juliana Osorio Morales \thanks{Instituto de Matem\'atica Luis Santal\'o (IMAS), UBA CONICET, Buenos Aires, Argentina
juli.osorio@gmail.com.} and Osvaldo P. Santill\'an \thanks{Instituto de Matem\'atica Luis Santal\'o (IMAS), UBA CONICET, Buenos Aires, Argentina
firenzecita@hotmail.com and osantil@dm.uba.ar.}}

\date {}
\maketitle

\begin{abstract}

The focus of the present work is on the Cauchy problem for the quadratic gravity models introduced in \cite{stelle}-\cite{stelle2}. These are renormalizable higher order derivative models of gravity, but at cost of ghostly states propagating in the phase space. A previous work on the subject is \cite{noakes}. The techniques employed here  differ slightly from those in \cite{noakes}, but the main conclusions agree. Furthermore, the analysis of the initial value formulation in \cite{noakes} is enlarged and the use of harmonic coordinates is clarified. In particular, it is shown that the initial constraints found \cite{noakes} include a redundant one. In other words, this constraint is satisfied when the equations of motion are taken into account. In addition, some terms that are not specified in \cite{noakes} are derived explicitly. This procedure facilitates application of some of the  mathematical theorems given in \cite{ringstrom}. As a consequence of these theorems, the existence of both $C^\infty$ solutions and maximal globally hyperbolic developments is proved. The obtained equations may be relevant for the stability analysis of the solutions under small perturbations of the initial data.
\end{abstract}

\section{Introduction}

The present work studies the Cauchy problem  for the quadratic gravity scenarios introduced by Stelle in  \cite{stelle}-\cite{stelle2}. The Stelle's equations of motion are of fourth order. A pioner work about the subject is \cite{noakes}, and the purpose of the present paper is to enlarge the results of that reference. There is a clear interest in the quantization of these theories and, for this reason, it is of fundamental importance to understand their Cauchy formulation first. 
 
A major feature of the quadratic gravity scenarios of \cite{stelle}-\cite{stelle2} is that they are renormalizable, which is one of the expected properties of a consistent quantum gravity model.  As it is well known, the quantization of GR yields a non renormalizable Quantum Field Theory. However, it is widely believed that a consistent quantum gravity theory should contain in the Lagrangian
terms with higher order derivatives of the metric. These terms are expected to play an insignificant role at low energies, but at high energy they may play a central role and  stabilize the divergent structure of the theory.

The reference \cite{pais} is a pioneer in the study of higher derivative theories in the context of Quantum Field Theory, and suggested that higher derivative terms may stabilize the divergent behavior of GR when the interaction with matter is turned on. Following these ideas, the works \cite{stelle}-\cite{stelle2} presented a concrete renormalizable gravity model including the quadratic terms $R^2$ and $R_{\mu\nu}R^{\mu\nu}$ in the Lagrangian. Later on, it was noticed that the Euclidean versions of these models are asymptotically free \cite{renor1}-\cite{renor4}, see also \cite{odintsov}.

The renormalization property of the higher derivative models found in \cite{stelle}-\cite{stelle2} is attractive from the theoretical point of view. However, its main problem comes from the so called Ostrogradski ghost \cite{ostrogradski}, which may be an unwanted feature in a consistent quantum gravity model. The reason for the appearance of a ghost is that the Hamiltonian is linear in the higher derivative fields momenta, and it is not bounded from below. This creates negative norm states when quantizing the theory. These modes propagate through the phase space and produce instabilities. In fact, when expanded around the flat Minkowski background the theory is renormalizable, but the kinetic terms of the  graviton and the massive spin two degree of freedom have opposite signs. This suggests that one of these states is a ghost.

Some references which attempt to avoid the problems described above are \cite{sala1}-\cite{sala4}. A specific prescription for removing ghosts in the theory was introduced in \cite{odintsov2}. Some attempts to avoid the ghost instabilities by reducing the phase space are described in \cite{chenlim2}. These ideas were pursued further in \cite{chenlim} and \cite{akitakob}. Additional issues related to unitarity in these models were studied in \cite{reno1}-\cite{reno3}. The use of the Stuckelberg trick in these scenarios for studying the high energy limit  in which the graviton mass tends to zero was discussed in \cite{hinterblicher}. 

There are also several classical aspects of these models which are of genuine interests. An example is the black hole physics that is predicted from these scenarios. Black hole solutions for a higher dimensional version of these models were considered in \cite{clasico1}-\cite{clasico2} and a precise numerical analysis of the asymptotic behaviour of these solutions was performed in \cite{clasico3}-\cite{clasico4}. The effect of the addition of a cosmological constant in the model was studied in \cite{clasico9}
-\cite{clasico10}. In addition, it was shown that a new branch of black hole solutions occurs along with the standard Schwarzschild
branch in these models. The standard and new branches cross at a point determined by a static
negative-eigenvalue eigenfunction of the Lichnerowicz operator. The role of these Lichnerowicz modes was reconsidered recently in \cite{lichnerowicz}.
The stability of black hole solutions mentioned above was studied in \cite{clasico3}-\cite{clasico7}. Furthermore, the first law of thermodynamics for the black holes arising Stelle's gravity was considered in \cite{clasico8}. Further aspects of interest from the physical point of view are described in \cite{salvio}.

In the present work, an important role is played by second order quasi-linear hyperbolic systems, whose generic form is given by
\begin{equation}\label{choquet}
g^{\mu\nu}(x, t, u_i)\frac{\partial u_q}{\partial x^\mu \partial x^\nu}=f_q(u_i,\partial u_i),
\end{equation}
where $u_q$ with $q=1,..,n$ is a vector constituted  the unknowns \cite{RelatividadChoquet}-\cite{foures}. Here the matrix $g^{pq}$ is the same for all the equations $q=1,..n$ and it is of normal hyperbolic type, that is
$g^{44}\leq  0$ and $g^{ij}x_i x_j$ is a positive definite form, with the latin indices indicating spatial directions. The explicit form of the non linearity $f_q(u_i,\partial_j u_i)$ is of interest because it may characterize some formal properties of the solutions of the model.

The structure of the present work is as follows. In Section 2 the main equations of the model are reviewed, and the initial value formulation is outlined.
In particular, the constraints to be satisfied for the initial conditions are worked out explicitly. In Section 3 the use of harmonic coordinates is clarified, and it is shown that they are locally consistent if some initial conditions are imposed. Due to the higher order nature of the Stelle's equations, the corresponding initial conditions contains two constraints which are not present in GR. The set of constraints found here agree with those in \cite{noakes}, although the techniques employed are not completely the same. On the other hand, it is shown that there is a redundant constraint in \cite{noakes}. In other words, it comes as a consequence of  the equations of motion. In Section 4 the degrees of freedom of the Stelle quadratic gravity are described. This material is not new, but it contains some intuitions that are helpful for understanding the procedure employed in Section 5. The reader interested in formal mathematical aspects may skip Section 4 and still be able to understand the main mathematical procedure employed in subsequent sections. In Section 5 the evolution equations of the Stelle gravity models are analysed, an an order reduction procedure bringing the system into the form (\ref{choquet}) is used. In these terms, the non linearity of the resulting system is characterized and, furthermore, it is shown that it satisfies a technical condition namely, {\it x-compactness}.
This condition permits to make statements about the mathematical nature of the solutions of the model.  The properties of these solutions are largely discussed in the conclusions given in Section 6. Some technical details employed throughout the text are outlined in the appendix A.

\section{The Stelle's equations and their initial value formulation}

\subsection{The main equations}
The action of the Stelle higher derivative gravity model is the following \cite{stelle}-\cite{stelle2}
\be\lb{bacho}
S=\int \bigg[\frac{1}{16 \pi G_N} R+\alpha  R_{\mu\nu}R^{\mu\nu}+\beta R^2\bigg]\sqrt{-g}\,d^4x+S_m.
\ee
It contains the Einstein term, proportional to the Ricci scalar $R$, plus two terms proportional to $ R_{\mu\nu}R^{\mu\nu}$
and $R^2$ whose roles are to stabilize the divergent behavior of the Einstein model.
Here $S_m$ is the matter Lagrangian and $\alpha$ and $\beta$ are parameters, whose values are fixed by the unknown physics at high energy scales. The equations of motion that are derived from this action are given by \cite{stelle}-\cite{stelle2}
\be\lb{ecmov} 
H_{\mu\nu}=\frac{1}{16 \pi G_N} G_{\mu\nu} + E_{\mu\nu}=\frac{1}{2}T_{\mu\nu}.
\ee
Here $G_{\mu\nu}$ is the standard Einstein tensor
\be\lb{einst}
G_{\mu\nu} = R_{\mu\nu} -\frac{1}{2} R g_{\mu\nu}, 
\ee
$T_{\mu\nu}$ is the energy momentum tensor for the matter fields and the quantity $E_{\mu\nu}$ is given by
$$
E_{\mu\nu} = (\alpha-2\beta)\nabla_\mu\nabla_\nu R-\alpha \square R_{\mu\nu}-(\frac{1}{2}\alpha-2\beta)g_{\mu\nu}\square R
+2\alpha R^{\alpha\beta}R_{\mu\alpha\nu\beta}
$$
\be\lb{e}
-2\beta R R_{\mu\nu}-\frac{1}{2}g_{\mu\nu}(\alpha R_{\alpha\beta}R^{\alpha\beta}-\beta R^2).
\ee
In addition, the identity $\nabla_\mu T^{\mu\nu}=0$ implies that 
$\nabla_\mu H^{\mu\nu}=0$. This is an important identity for proving that harmonic coordinates are consistent, as it will be discussed below.  The present analysis is focused on the vacuum case $T_{\mu\nu}=0$, although several aspects may be generalized when matter fields are turned on.

   It will be  useful to express the equations of motion (\ref{ecmov}) in several equivalent forms. 
First of all, it is not difficult to prove that  equations (\ref{ecmov}) may be expressed as follows
$$
 \frac{1}{16 \pi G_N}(R_{\mu\nu} -\frac{1}{2} R g_{\mu\nu}) - 2\beta R(R_{\mu\nu}-\frac{1}{2}R g_{\mu\nu})
+(2\beta-\alpha) (g_{\mu\nu} \square -\nabla_\mu\nabla_\nu) R
$$
\be\lb{gorila2}
+ \alpha\square(R_{\mu\nu}-\frac{1}{2} R \, g_{\mu\nu})
+
2\alpha (R_{\mu\rho\nu\sigma} -\frac{1}{4} g_{\mu\nu}\, R_{\rho\sigma}) (R^{\rho\sigma}-\frac{1}{2}g^{\rho\sigma}R)
\ee
$$
+\alpha (R_{\mu\nu} -\frac{1}{2} g_{\mu\nu}\, R)R+\frac{1}{4}(\alpha-4\beta) g_{\mu\nu}R^2=0.
$$
It may be even convenient to write the last expression in terms of the Einstein tensor (\ref{einst}) as much as possible. A convenient expression is 
$$
 \alpha\square G_{\mu\nu} +\frac{1}{16 \pi G_N}G_{\mu\nu} +(\alpha- 2\beta) RG_{\mu\nu}
+(2\beta-\alpha) (g_{\mu\nu} \square -\nabla_\mu\nabla_\nu) R
$$
\be\lb{gorila4}
+2\alpha (R_{\mu\rho\nu\sigma} -\frac{1}{4} g_{\mu\nu}\, G_{\rho\sigma}) G^{\rho\sigma}+\frac{1}{2}(\alpha-2\beta) g_{\mu\nu}R^2=0.
\ee
As it is well known, the Einstein tensor $G_{\mu\nu}$ is a expression of second order in terms of the metric $g_{\mu\nu}$. Thus, the equations (\ref{gorila4}) are of fourth order in the unknowns $g_{\mu\nu}$.
\subsection{The initial value formulation of the model}
In standard GR, some of the Einstein equations, when projected over an initial spatial surface, become of first order and are interpreted as constraints for the initial data.  It is important to identify the initial constraints for the Stelle gravity model (\ref{bacho}). These are by definition the equations of motion with order strictly less than four, projected on an initial Cauchy hypersurface.

Assume that a globally hyperbolic solution
($M$, $g$) of the Stelle's equations (\ref{gorila4}) has been constructed. Then the space-time $M$ can be foliated by spatial hypersurfaces $\Sigma_t$ parametrized by a global time function $t$, whose gradient is never vanishing  \cite{RelatividadChoquet}-\cite{O Neill}. Let $n_a$ be a unit vector orthogonal to the hypersurfaces $\Sigma_t$. Then the metric $g_{\mu\nu}$ induces a spatial metric $h_{\mu\nu}$ in $\Sigma_t$ given by 
$$
h_{\mu\nu}=g_{\mu\nu}+n_\mu\otimes n_\nu.
$$
The vector $t^\mu$ defined by the condition $t^\mu \nabla_\mu t=1$ represents the flow of time $t$ through the space-time ($M$, $g$). Its spatial and time components are 
$$
 N_\mu=h_{\mu\nu}t^\nu,\qquad N=-t^\mu n_\mu,
$$
respectively. The quantity $N_\mu$ is known as the shift vector and $N$ as the lapse function. 
Given a generic vector $A^\mu$, it can be decomposed as $A^\mu=n^\mu A_o+A_{t}$, where the first part is orthogonal to $\Sigma$ and $A_t$ is the tangent part. The quantity $h_\mu^\nu$ is a projector over the tangent space $T\Sigma$, that is, $h^\mu_\nu A^\nu=A_{t}^\mu$. An analogous formula holds for tensor fields $A_{\mu_1...\mu_k}^{\;\;\;\;\;\;\;\;\;\nu_1..\nu_l}$. As it is well known, the projection 
$G_{\mu\nu}n^\nu|_{t=0}$ is of first order in time derivatives \cite{Wald}. More precisely, the spatial and time components of this projection at $\Sigma$ are given by
\be\lb{cod}
G_{\mu\nu}n^\mu h^{\nu}_\alpha=D_{\mu}k^{\mu}_{\alpha}-D_{\alpha}k^{\mu}_{\mu},\qquad
G_{\mu\nu}n^\mu n^\nu=\frac{1}{2}[R^{(3)}+(k_{\mu}^\mu)^2-k_{\mu\nu}k^{\mu\nu}].
\ee
Here the following quantity 
$$
k_{\mu\nu}=\frac{1}{2}{\cal L}_n h_{\mu\nu},
$$
has been introduced, with ${\cal L}_n$ the standard Lie derivative along the $n$ direction. In addition $R^{(3)}$ is the curvature corresponding to $h_{\mu\nu}$ and $D_\alpha$ is the corresponding three dimensional covariant derivative. These objects are defined on the tangent space $T\Sigma$ of the surface $\Sigma$ corresponding to the time $t=0$, and are of first order with respect to the time derivative $\partial_t$. Details of these assertions can be found in the standard textbooks \cite{RelatividadChoquet}-\cite{O Neill}, or in the extensive reference \cite{gourgu}. Given these expressions, it is tempting to examine the projection of (\ref{gorila4}) on the customary directions of GR.  Consider for instance the Stelle's equations projected on the $nn$ directions 
$$
 \alpha n^\mu n^\nu \square G_{\mu\nu} +\frac{1}{16 \pi G_N}n^\mu n^\nu G_{\mu\nu} +(\alpha- 2\beta)  n^\mu n^\nu G_{\mu\nu}R
+(2\beta-\alpha) n^\mu n^\nu(g_{\mu\nu} \square -\nabla_\mu\nabla_\nu) R
$$
\be\lb{gorilap4}
+2\alpha (n^\mu n^\nu R_{\mu\rho\nu\sigma} -\frac{1}{4} n^\mu n^\nu g_{\mu\nu}\, G_{\rho\sigma}) G^{\rho\sigma}+\frac{1}{2}(\alpha-2\beta) n^\mu n^\nu g_{\mu\nu}R^2=0.
\ee
As $G_{\mu\nu}$ and $R$ are expressions with at most two derivatives, the only terms that may be of fourth order are those related to the D'Alambertian $\square$ or to the covariant derivatives $\nabla_\alpha$. However, from the identity
$$
\nabla_\alpha \nabla_\beta (G_{\mu\nu}n^\mu n^\nu)=n^\mu n^\nu\nabla_\alpha \nabla_\beta G_{\mu\nu}+(n^\nu\nabla_\alpha n^\mu+n^\nu\nabla_\alpha n^\mu) \nabla_\beta G_{\mu\nu}+(n^\nu\nabla_\beta n^\mu+n^\nu\nabla_\beta n^\mu) \nabla_\alpha G_{\mu\nu}
$$
$$
+(n^\mu \nabla_\alpha\nabla_\beta n^\nu+\nabla_\alpha n^\mu \nabla_\beta n^\nu+\nabla_\alpha n^\nu \nabla_\beta n^\mu+n^\nu \nabla_\alpha\nabla_\beta n^\mu)G_{\mu\nu},
$$
it can be deduced that
$$
n^\mu n^\nu\square G_{\mu\nu}=\square G_{\mu\nu}n^\mu n^\nu-g^{\alpha\beta}(n^\nu\nabla_\alpha n^\mu+n^\nu\nabla_\alpha n^\mu) \nabla_\beta G_{\mu\nu}-g^{\alpha\beta}(n^\nu\nabla_\beta n^\mu+n^\nu\nabla_\beta n^\mu) \nabla_\alpha G_{\mu\nu}
$$
\be\lb{riga}
-(n^\mu \square n^\nu+g^{\alpha\beta}\nabla_\alpha n^\mu \nabla_\beta n^\nu+g^{\alpha\beta}\nabla_\alpha n^\nu \nabla_\beta n^\mu+n^\nu \square n^\mu)G_{\mu\nu}.
\ee
Since $G_{\mu\nu}n^\mu n^\nu$ contains no second time derivatives, the first term in the left contains at most third time derivatives. The other terms are clearly also of order less than four in time derivatives.
In addition
$$
(n^\nu n^\mu g_{\mu\nu} \square -n^\mu n^\nu \nabla_\mu\nabla_\nu) R=(\square-n^\mu n^\nu \nabla_\mu\nabla_\nu) R
$$
\be\lb{riga2}
=[(h^{\alpha\beta}+n^\alpha n^\beta)\nabla_\alpha \nabla_\beta -n^\mu n^\nu \nabla_\mu\nabla_\nu] R=h^{\alpha\beta}\nabla_\alpha \nabla_\beta R=0.
\ee
The last formula clearly does not contains second time derivatives of $R$ and therefore is at most of third order. From (\ref{riga})-(\ref{riga2}) it follows that (\ref{gorilap4}) is a expression of third order with respect to time derivatives and therefore, when projected on the initial Cauchy surface $\Sigma$, it becomes a constraint.

Consider now the projection $nh$
$$
 \alpha n^\mu h^\nu_\alpha \square G_{\mu\nu} +\frac{1}{16 \pi G_N}n^\mu h^\nu_\alpha G_{\mu\nu} +(\alpha- 2\beta)  n^\mu h^\nu_\alpha G_{\mu\nu}R
+(2\beta-\alpha) n^\mu h^\nu_\alpha(g_{\mu\nu} \square -\nabla_\mu\nabla_\nu) R
$$
\be\lb{gorilappp4}
+2\alpha (n^\mu h^\nu_\alpha R_{\mu\rho\nu\sigma} -\frac{1}{4} n^\mu h^\nu_\alpha g_{\mu\nu}\, G_{\rho\sigma}) G^{\rho\sigma}+\frac{1}{2}(\alpha-2\beta) n^\mu h^\nu_\alpha g_{\mu\nu}R^2=0.
\ee
By use of an argument similar to the one leading to (\ref{riga}) it can be deduced that
$$
n^\mu h^\nu_\alpha\square G_{\mu\nu}=\square G_{\mu\nu}n^\mu h^\nu_\alpha-g^{\gamma\beta}(n^\mu\nabla_\gamma h^\nu_\alpha+h^\nu_\alpha\nabla_\gamma n^\mu) \nabla_\beta G_{\mu\nu}-g^{\gamma\beta}(h^\nu_\alpha\nabla_\beta n^\mu+h^\nu_\alpha\nabla_\beta n^\mu) \nabla_\gamma G_{\mu\nu}
$$
\be\lb{rigas}
-(n^\mu \square h^\nu_\alpha +g^{\alpha\beta}\nabla_\alpha n^\mu \nabla_\beta h^\nu_\alpha+g^{\alpha\beta}\nabla_\alpha h^\nu_\alpha \nabla_\beta n^\mu+h^\nu_\alpha \square n^\mu)G_{\mu\nu}.
\ee
By taking into account that $n^\mu h^\nu_\alpha G_{\mu\nu}$ is a expression involving first order time derivatives, it follows from (\ref{rigas}) that $n^\mu h^\nu_\alpha\square G_{\mu\nu}$ is at most of third order. In addition, as $h_\mu^\alpha g_{\alpha\nu}=h_{\mu\nu}$ and $n^\mu h_{\mu\nu}=0$ by the orthogonality condition, it follows that
$$
n^\mu h^\nu_\alpha(g_{\mu\nu} \square -\nabla_\mu\nabla_\nu) R=-n^\mu h^\nu_\alpha \nabla_\nu\nabla_\mu R=n^\mu D^{(3)}_\alpha \nabla_\mu R,
$$
where in the last step, the fact that $\nabla_\mu \nabla_\nu=\nabla_\nu\nabla_\mu$ when acting on scalar functions has been taken into account. The operator $D^{(3)}_\alpha$
contains no time derivatives and therefore the last is a expression at most of third order in time derivatives. Thus (\ref{gorilappp4}) is also a constraint. Its explicit form is
$$
 \alpha n^\mu h^\nu_\alpha \square G_{\mu\nu} +\frac{1}{16 \pi G_N}n^\mu h^\nu_\alpha G_{\mu\nu} +(\alpha- 2\beta)  n^\mu h^\nu_\alpha G_{\mu\nu}R
$$ 
 \be\lb{gorilapp4}
-(2\beta-\alpha) n^\mu D^{(3)}_\alpha \nabla_\nu R
+2\alpha n^\mu h^\nu_\alpha R_{\mu\rho\nu\sigma}  G^{\rho\sigma}=0.
\ee
From all this discussion, it follows that the set initial conditions for the Stelle's equations (\ref{gorila4}) is composed as follows. First, define a spatial hypersurface $\Sigma$ corresponding to the time $t=0$. On this hypersurface, introduce an initial metric $g_{(0)\mu\nu}=h_{(0)\mu\nu}+n_{(0)\mu}\otimes n_{(0)\nu}$ together with three symmetric quantities $k_{(0)\mu\nu}$, $G_{(0)\mu\nu}$ and $K_{(0)\mu\nu}$. The initial conditions then are given by \footnote{The last condition may be replaced by ${\cal L}_n G_{\mu\nu}|_{t=0}=K'_{(0)\mu\nu}$. But from the known expression ${\cal L}_n T_{\mu_1..\mu_k}^{\nu_1..\nu_2}=n^\mu \nabla_\nu T_{\mu_1..\mu_k}^{\nu_1..\nu_l}+\sum_{i=1}^k T_{\mu_1..\sigma..\mu_k}^{\nu_1..\nu_l}\nabla_{\mu_i}n^\sigma-\sum_{i=1}^k T_{\mu_1..\mu_k}^{\nu_1..\sigma..\nu_l}\nabla_{\sigma}n^{\mu_i}$, it follows that both data give the same information.}
$$
g_{\mu\nu}|_{t=0}=g_{(0)\mu\nu}, \qquad {\cal L}_n h_{\mu\nu}|_{t=0}=k_{\mu\nu},
$$
\be\lb{initial}
G_{\mu\nu}|_{t=0}=G_{(0)\mu\nu},\qquad n^\alpha \nabla_\alpha G_{\mu\nu}|_{t=0}=K_{(0)\mu\nu}.
\ee
The first two formulas in (\ref{initial}) are present in GR, the last two are new and define the second and the third time derivatives of the metric $g_{\mu\nu}$ respectively. Note that the value of
$R$ on $\Sigma$ is defined by this information, since $R$ is proportional to the trace of $G_{\mu\nu}$. The quantities $g_{(0)\mu\nu}=h_{(0)\mu\nu}+n_{(0)\mu}\otimes n_{(0)\nu}$, $k_{(0)\mu\nu}$, $G_{(0)\mu\nu}$ and $K_{(0)\mu\nu}$ are not arbitrary, but related to each other by the constraints (\ref{gorilap4}) and (\ref{gorilapp4}). These quantities  are the initial data for constructing the space-time evolution $(M,g)$.

\section{The Stelle's equations in harmonic coordinates}
In the previous section, it has been shown that the dynamical Stelle's equations are the six spatial components of (\ref{gorila4}). The remaining equations instead are simply constraints. The unknowns are the ten components $g_{\mu\nu}$ of the metric, which shows that the system is overdetermined. This reflects the invariance of the model under diffeomorphisms. In order to remove this ambiguity, a coordinate  gauge should be imposed. A  gauge that is successful in GR is the harmonic one.
Its advantage is that, in these coordinates, the Einstein equations take the form (\ref{choquet}). Of course, this conclusion does not follow
directly for the Stelle's quadratic gravity, since the equations are of higher order. However,  by the use of these coordinates and by  use of the known procedure of order reduction, a quasi-linear hyperbolic second order system equivalent to (\ref{choquet}) can be constructed. The strategy is then to characterize the solutions of Stelle's equations from the general properties of those systems.

\subsection{The consistency of the use of harmonic coordinates}

The harmonic gauge is simple to describe. For an arbitrary space-time ($M$, $g_{\mu\nu}$) locally parametrized by some coordinates $x^\mu$, the Ricci tensor is given by the general formula
\begin{equation}\label{desco2}
R_{\mu\nu}=-\frac{1}{2}g^{\alpha\beta}\partial_{\alpha}\partial_{\beta}g_{\mu\nu}+Q_{\mu\nu}(g,\partial g)+\frac{1}{2}(
g_{\mu\beta}\partial_\nu F^\beta+g_{\nu\beta}\partial_\mu F^\beta),
\end{equation} 
where $Q_{\mu\nu}(g,\partial g)$ is a quantity which depends on the metric and its first derivatives. Its explicit form is
$$
Q^{\mu\nu}=g^{\alpha\beta}[\Gamma_{\alpha\gamma}^{\mu}\partial_{\beta} g^{\nu\gamma}+\Gamma_{\alpha\gamma}^{\nu}\partial_{\beta} g^{\mu\gamma}-2\Gamma_{\alpha\beta}^{\gamma}\partial_{\gamma} g^{\nu\mu}].
$$
In addition
\be\lb{tearmo}
F^\alpha=g^{\mu\nu}\Gamma_{\mu\nu}^\alpha=\frac{1}{\sqrt{-g}}\frac{\partial}{\partial x^\alpha}\bigg(\sqrt{-g}g^{\alpha\beta}\bigg),\qquad \alpha=1,\;2,\;3,\;4.
\ee
The harmonic conditions is by definition $F^\alpha=0$, and the harmonic coordinates are those which satisfy it. This is the gauge to be employed in the following. It is clear from the previous formulas that the Ricci tensor, the scalar curvature and the Einstein tensor in harmonic coordinates are given by
$$
R^F_{\mu\nu}=-\frac{1}{2}g^{\alpha\beta}\partial_{\alpha}\partial_{\beta}g_{\mu\nu}+Q_{\mu\nu}(g,\partial g),\qquad
R^F=-\frac{1}{2}g^{\alpha\beta}g^{\sigma\rho}\partial_{\alpha}\partial_{\beta}g_{\sigma\rho}+Q(g,\partial g).
$$
\be\lb{armonia}
G^F_{\mu\nu}=-\frac{1}{2}g^{\alpha\beta}\partial_{\alpha}\partial_{\beta}g_{\mu\nu}+\frac{1}{4}g_{\mu\nu}g^{\alpha\beta}g^{\sigma\rho}\partial_{\alpha}\partial_{\beta}g_{\sigma\rho}+Q_{\mu\nu}(g,\partial g)-\frac{1}{2}g_{\mu\nu}Q(g,\partial g),
\ee
respectively. By taking the last formulas as the definitions of $R^F_{\mu\nu}$, $R^F$ and $G_{\mu\nu}^F$ it follows that
\be\lb{desco}
R^F_{\mu\nu}=R_{\mu\nu}-\frac{1}{2}(
g_{\mu\beta}\partial_\nu F^\beta+g_{\nu\beta}\partial_\mu F^\beta),\qquad
R^F=R-\partial_\alpha F^\alpha
\ee
\be\lb{armonia2}
G^F_{\mu\nu}=G_{\mu\nu}-\frac{1}{2}(
g_{\mu\beta}\partial_\nu F^\beta+g_{\nu\beta}\partial_\mu F^\beta-g_{\mu\nu}\partial_\alpha F^\alpha).
\ee
The previous discussion shows that the Ricci tensor in harmonic coordinates ($F^\alpha=0$) becomes a quasi-diagonal second-order operator for the components of $g$, since it has the form $2R_{\mu\nu}= -g^{\alpha\beta}\partial_{\alpha}\partial_{\beta}g_{\mu\nu}+2Q_{\mu\nu}$, where the last term contains only first order terms. If the harmonic condition $F^\alpha=0$ is not fulfilled, then the expression for $R_{\mu\nu}$  is not quasi-diagonal and the  techniques derived from (\ref{choquet}) should not be applied.

The description given  above suggests that the condition $F^\alpha=0$ may be of practical convenience. However, it may not be legitimate to assume that the harmonic condition $F^\alpha=0$ holds, even if it is satisfied on certain subset $\Omega$ the initial Cauchy surface $\Sigma$. The problem is that, given solution $g_{\mu\nu}$ of (\ref{gorila4}), it may be the case that $F^\alpha\neq 0$  in the Cauchy development $D(\Omega)$, even if initially $F^\alpha|_{t=0}=0$ on $\Omega$. If this is so, then the choice $F^\alpha=0$ is inconsistent. Thus, the equations that describe the evolution of the quantity $F^\alpha$ should be obtained in order to check that, given suitable initial conditions, the solution is $F^\alpha=0$ for all the times $t\geq 0$.

The evolution equations for $F^\alpha$ are obtained as follows. The Stelle equations (\ref{ecmov}) in any coordinate system are given by
$$
H_{\mu\nu}= G_{\mu\nu}  +(\alpha-2\beta)RG_{\mu\nu}
+(2\beta-\alpha) (g_{\mu\nu} \square -\nabla_\mu\nabla_\nu) R
$$
\be\lb{gorila6}
+ \alpha\square G_{\mu\nu}
+
\bigg[2\alpha R_{\mu\rho\nu\sigma} -\frac{\alpha}{2} g_{\mu\nu} G_{\sigma\rho} 
\bigg] G^{\sigma\rho}+\frac{1}{2}(\alpha-2\beta) g_{\mu\nu}R^2=T_{\mu\nu},
\ee
while the use of harmonic gauge brings them to the form
$$
 H_{\mu\nu}^F=G^F_{\mu\nu}  + (\alpha-2\beta)R^F G^F_{\mu\nu}
+(2\beta-\alpha) (g_{\mu\nu} \square^F -\nabla_\mu\nabla_\nu) R^F
$$
\be\lb{gorila7}
+ \alpha\square^F G^F_{\mu\nu}
+
\bigg[2\alpha R_{\mu\rho\nu\sigma} -\frac{\alpha}{2} g_{\mu\nu} G^F_{\sigma\rho} 
\bigg] G^{F\sigma\rho}+\frac{1}{2}(\alpha-2\beta) g_{\mu\nu}(R^F)^2=T_{\mu\nu},.
\ee
Here the notation $\square^F$ requires a short explanation. The laplacian acting on any scalar function, in particular on $R$, is given in local coordinates by
$$
\square R= g^{\alpha\beta}\partial_\alpha\partial_\beta R+F^\alpha R.
$$
On the other hand if the harmonic coordinate condition (\ref{tearmo}) is imposed, then the second term is zero. Thus, the simple formula
$$
\square^F R^F=g^{\alpha\beta}\partial_\alpha\partial_\beta R^F,
$$
is obtained in this gauge. 
Analogous considerations follow for $\square^F G^F_{\mu\nu}$. In this situation, the action of the D'Alambertian on a tensor like $G_{\mu\nu}$ is slightly more complicated
than for scalar fields. Nevertheless, an inspection of the relevant formulas shows that, even for this situation, $\square^F T_{\mu\nu}=\square T_{\mu\nu} - F^\alpha \partial_\alpha T_{\mu\nu}$.

Assume now that a particular solution $g_{\mu\nu}$ of (\ref{gorila7}) has been found. The tensor $H_{\mu\nu}$ is divergence free, this is a geometrical identity, the analogous of $\nabla^\mu G_{\mu\nu}=0$ in GR for the present model. The energy momentum tensor $T_{\mu\nu}$ is divergence free (in particular, the  tensor $T_{\mu\nu}=0$ has zero divergence). Therefore $\nabla^\mu H_{\mu\nu}=0$ and, from (\ref{gorila7}), it also follows that $\nabla^\mu H_{\mu\nu}^F=0$. The difference therefore must satisfy $\nabla^\mu (H_{\mu\nu}-H^F_{\mu\nu})=0$.
The explicit expression for this difference is 
$$
\delta H_{\mu\nu}=H_{\mu\nu}-H^F_{\mu\nu}=\delta G_{\mu\nu}+ \alpha\square \delta G_{\mu\nu}-\alpha F^\delta \partial_\delta G_{\mu\nu}^F+(2\beta-\alpha)(\partial_\alpha F^\alpha G_{\mu\nu}+R^F \delta G_{\mu\nu}+\partial_\alpha F^\alpha \delta G_{\mu\nu})
$$
$$
+\frac{1}{2}(\alpha-2\beta) g_{\mu\nu}(2R^F\partial_\alpha F^\alpha+\partial_\alpha F^\alpha \partial_\beta F^\beta)
+\bigg[2\alpha R_{\mu\rho\nu\sigma} -\frac{\alpha}{2} g_{\mu\nu} G^F_{\sigma\rho} 
\bigg] \delta G^{\sigma\rho}-\frac{\alpha}{2}g_{\mu\nu}\delta G_{\sigma\rho}G^{F\sigma\rho}
$$
\be\lb{difer}
-\frac{\alpha}{2}g_{\mu\nu}\delta G_{\sigma\rho}\delta G^{\sigma\rho}+(2\beta-\alpha)(g_{\mu\nu}\square-\nabla_\mu\nabla_\nu)(\partial_\alpha F^\alpha)+(2\beta-\alpha)g_{\mu\nu}F^\alpha \partial_\alpha R^F.
\ee
The expression for $\delta G_{\mu\nu}$ in terms of $F^\alpha$ can be read off from (\ref{armonia2}), the result is
\be\lb{gang}
\delta G_{\mu\nu}=\frac{1}{2}(
g_{\mu\beta}\partial_\nu F^\beta+g_{\nu\beta}\partial_\mu F^\beta-g_{\mu\nu}\partial_\alpha F^\alpha).
\ee
If the last definition is introduced into the expression for $\delta H_{\mu\nu}$  derived above then, after imposing that $\nabla^\mu \delta H_{\mu\nu}=0$,
a fourth order term $\square \nabla_\mu \delta G^\mu_{\nu}$ will appear.  This term is induced by the divergence of the second term in the right hand side of (\ref{difer}). A fourth order non linear equation is difficult to deal with.

An approach to sort out these problems out is to add new variables, and to construct  a second order system\footnote{Note that this system is just a simplified version of (\ref{choquet}), in the sense that the quantity $g^{\mu\nu}(x^\alpha)$ does not depend on the unknowns $\eta_\mu$ or $\partial_\mu\eta_\nu$ while in (\ref{choquet}) $g^{\mu\nu}=g^{\mu\nu}(x^\alpha, u_q)$.} 
\be\lb{prat}
g^{\mu\nu}(x^\alpha)\frac{\partial^2 \eta_q}{\partial x^\mu\partial x^\nu}=f_q(\eta_l,\partial_\mu \eta_l),
\ee
completely equivalent to the equations  $\nabla^\mu \delta H_{\mu\nu}=0$ obtained from (\ref{difer}).
Here $\eta_\alpha$ are the unknowns and the second derivatives $\partial_\mu \partial_\nu \eta_q$ are allowed only on the right hand. Furthermore the quantities 
$g^{\mu\nu}(x^\alpha)$ should not contain the unknowns.
It is also desirable that the resulting non linearity $f_q(\eta_l,\partial_\mu \eta_l)$ is such that $f_q(0,0)=0$. If this property is fulfilled then, by imposing the initial conditions $\eta_\alpha=\partial_\beta \eta_\alpha=0$, it follows from (\ref{prat}) that the second derivatives $\partial_{\alpha}\partial_\beta \eta_\nu$ will also be zero. If the non linearity $f_q(\eta_l,\partial_\mu \eta_l)$ is suitable enough, then by taking further derivatives in (\ref{prat}) it follows that all the derivatives  of $\eta_\alpha$ are zero and and in particular, that $F^\mu=0$.  

The order reduction procedure sketched in the  previous paragraph is implemented as follows. First, note that the fourth order contribution to $\nabla^\mu \delta H_{\mu\nu}$  comes from the term  $\square \nabla_\mu \delta G^\mu_{\nu}$. The fact that this is a fourth order term follows from the definition (\ref{gang}), which shows that $\delta G^\mu_{\nu}$ is a first order quantity in $F^\mu$.  This suggests that the system can be converted into a second order one such as (\ref{prat}) if the second order expression $\nabla_\mu \delta G^\mu_{\nu}$ is considered as an independent variable. The independent variables are then $F^\mu$ and $\nabla_\mu \delta G^\mu_{\nu}$. The system should then be supplemented with a second order equation for $F^\alpha$. This is obtained by taking the divergence of $\delta G^\mu_{\nu}$ in (\ref{gang}) and adding the resulting equations to the system (\ref{difer}). This divergence is explicitly
\be\lb{id12}
g^{\mu\nu}\partial_{\mu}\partial_{\nu}F^\alpha=A^{\alpha\beta}_\gamma \partial_\beta F^\gamma+g^{\alpha\nu}\nabla_\mu \delta G^\mu_{\nu}.
\ee
Here the quantities $A^{\alpha\beta}_\gamma$ are local functions of the space-time coordinates, their expressions will not be important in the following.
On the other hand, the condition $\nabla^\mu \delta H_{\mu\nu}=0$ should be expressed in terms of $F^\alpha$ and $\partial_\mu F^\alpha$ everywhere by use of (\ref{armonia2}), with the exception for the terms $\nabla_\mu \delta G^\mu_{\nu}$. The last are considered as independent unknowns. These replacements bring (\ref{difer}) into the following form
$$
 \alpha\square\nabla_\mu (\delta G_{\nu}^\mu)=L_\nu(R^\alpha_{\beta\gamma\delta}, \partial_\alpha F^\beta, \partial_\alpha\partial_\beta F^\gamma)-\nabla_\mu (\delta G_{\nu}^\mu)-(2\beta-\alpha)\bigg[G_\nu^{\;\;\beta} \partial_\alpha\partial_\beta F^\alpha + \delta G_\nu^{\;\;\beta}\,\partial_\beta R^F
$$
$$
+
R^F \nabla_\mu (\delta G_{\nu}^\mu)+\delta G_\nu^{\;\;\beta}\partial_\alpha \partial_\beta F^\alpha 
+(\partial_\alpha  F^\alpha )\nabla_\mu (\delta G_{\nu}^\mu)\bigg]
-(\alpha-2\beta)(R^F\partial_\alpha \partial_\nu F^\alpha +\partial_\alpha F^\alpha \,\,\partial_\nu R^F+\partial_\beta F^\beta\,\partial_\alpha \partial_\nu F^\alpha )
$$
$$
- \delta G^{\sigma\rho}\nabla^\mu \bigg[2\alpha R_{\mu\rho\nu\sigma} -\frac{\alpha}{2} g_{\mu\nu} G^F_{\sigma\rho} 
\bigg]+ \bigg[2\alpha R_{\mu\rho\nu\sigma} -\frac{\alpha}{2} g_{\mu\nu} G^F_{\sigma\rho} 
\bigg] \nabla^\mu \delta G^{\sigma\rho}+\frac{\alpha}{2}G^{F\sigma\rho}\,\nabla_\nu \delta G_{\sigma\rho}+\frac{\alpha}{2}\delta G_{\sigma\rho}\,\nabla_\nu G^{F\sigma\rho}
$$
\be\lb{block}
+\alpha \delta G_{\sigma\rho}\nabla_\nu \delta G^{\sigma\rho}-(2\beta-\alpha)R^\delta_{\;\;\nu} \partial_\delta \partial_\alpha F^\alpha+(2\beta-\alpha)(\partial_\nu F^\alpha) \partial_\alpha R^F-(2\beta-\alpha) F^\alpha \nabla_\nu \partial_\alpha R^F.
\ee
In obtaining these expressions, the fact that $\nabla^\mu \square G_{\mu\nu}- \square \nabla^\mu G_{\mu\nu}\neq0$ was taken into account. This difference was denoted  as $L_\nu(R^\alpha_{\beta\gamma\delta}, \partial_\alpha F^\beta, \partial_\alpha\partial_\beta F^\gamma)$ in (\ref{block}), and is a linear combination in both $\partial_\alpha F^\beta$ and $\partial_\alpha\partial_\beta F^\gamma$. The explicit form of these terms will  not be relevant in the following discussion, but it is important to remark that $L_\nu(R^\alpha_{\beta\gamma\delta}, 0, 0)=0$.

The system composed by the equations (\ref{id12})-(\ref{block}) is of second order in the variables $\nabla_\mu \delta G^\mu_\nu$ and $F^\alpha$, which is a  desired feature. However, the sough-for system (\ref{prat}) contains second derivatives only on the left, and these second derivatives are multiplied by quantities $g^{\mu\nu}(x^\alpha)$ which do not depend on the unknowns $\eta_\alpha$. On the other hand, in the obtained equations (\ref{id12})-(\ref{block}) there are still terms such as $\partial_\beta F^\beta\,\partial_\alpha \partial_\nu F^\alpha$ which are not compatible with the form (\ref{prat}).  This problem can be avoided by adding the partial derivatives $\Phi_\beta^\alpha=\partial_\beta F^\alpha$ as a new set of variables. Take the partial derivatives $\partial_\beta$  of (\ref{id12}), with $\beta=1,..,4$, and add the resulting equations  to the system (\ref{id12})-(\ref{block}). The resulting equations are now
\be\lb{idol}
g^{\mu\nu}\partial_{\mu}\partial_{\nu}F^\alpha=-A^{\alpha\beta}_\gamma \partial_\beta F^\gamma+g^{\alpha\nu}\nabla_\mu \delta G^\mu_{\nu},
\ee
\be\lb{idol2}
g^{\mu\nu}\partial_{\mu}\partial_{\nu}\Phi_\gamma^\alpha=-(\partial_\gamma g^{\mu\nu})\partial_{\mu}\Phi_\nu^\alpha-A^{\alpha\beta}_\delta \partial_\beta \Phi^\delta_\gamma-A^{\alpha\beta}_{\delta\gamma} \Phi_\beta^\gamma+(\partial_\gamma g^{\alpha\nu})\nabla_\mu \delta G^\mu_{\nu}+g^{\alpha\nu}\partial_\gamma (\nabla_\mu \delta G^\mu_{\nu}),
\ee
$$
 \alpha\square\nabla_\mu (\delta G_{\nu}^\mu)=-L_\nu(R^\alpha_{\beta\gamma\delta}, \Phi_\alpha^\beta, \partial_\alpha\Phi_\beta^\gamma)-\nabla_\mu (\delta G_{\nu}^\mu)-(2\beta-\alpha)\bigg[G_\nu^{\;\;\beta} \partial_\alpha\Phi_\beta^\alpha + \delta G_\nu^{\;\;\beta}\,\partial_\beta R^F
$$
$$
+
R^F \nabla_\mu (\delta G_{\nu}^\mu)+\delta G_\nu^{\;\;\beta}\partial_\alpha \Phi_\beta^\alpha+\Phi_\alpha^\alpha\nabla_\mu (\delta G_{\nu}^\mu)\bigg]
-(\alpha-2\beta)(R^F\partial_\alpha \Phi_\nu^\alpha +\Phi_\alpha^\alpha \,\,\partial_\nu R^F+\Phi_\beta^\beta\,\partial_\alpha \Phi_\nu^\alpha ) 
$$
$$
- \delta G^{\sigma\rho}\nabla^\mu \bigg[2\alpha R_{\mu\rho\nu\sigma} -\frac{\alpha}{2} g_{\mu\nu} G^F_{\sigma\rho}\bigg]- \bigg[2\alpha R_{\mu\rho\nu\sigma} -\frac{\alpha}{2} g_{\mu\nu} G^F_{\sigma\rho} 
\bigg] \nabla^\mu \delta G^{\sigma\rho}+\frac{\alpha}{2}G^{F\sigma\rho}\,\nabla_\nu \delta G_{\sigma\rho}
$$
\be\lb{block2}
+\frac{\alpha}{2}\delta G_{\sigma\rho}\,\nabla_\nu G^{F\sigma\rho}+\alpha \delta G_{\sigma\rho}\nabla_\nu \delta G^{\sigma\rho}-(2\beta-\alpha)[R^\delta_{\;\;\nu} \partial_\delta \Phi^\alpha_{\alpha}+\Phi_\nu^\alpha \partial_\alpha R^F+ F^\alpha \nabla_\nu \partial_\alpha R^F].
\ee
As before,  expressions such as $\nabla_\mu \delta G_{\rho\sigma}$ or $\delta G_\rho^\sigma$ in  (\ref{block}) should be written  in terms of $F^\alpha$ and $\Phi_\alpha^\beta$, except for the variables $\nabla_\mu (\delta G_{\nu}^\mu)$. In these terms, define the vector composed by the unknowns
\be\lb{peta}
\eta_\alpha=(F^\beta, \Phi_\gamma^\beta, \nabla_\mu \delta G^\mu_\nu).
\ee
Then it is seen that the system (\ref{idol})-(\ref{block2}) obtained above is of the desired form  (\ref{prat}). The non linearity $f_q(\eta_\alpha,\partial_\mu \eta_\alpha)$ can be read from the right hand of the formulas (\ref{idol})-(\ref{block2}). For instance, from (\ref{idol}) it is seen that $$f^\alpha_1(\eta_\alpha,\partial_\mu \eta_\alpha)=-A^{\alpha\beta}_\gamma \partial_\beta F^\gamma+g^{\alpha\nu}\nabla_\mu \delta G^\mu_{\nu},$$ and that $f^\alpha_1(0,0)=0$. The same argument follows for the other non linearities, although their expressions are a bit more cumbersome. The quantity $g^{\mu\nu}$ in the left hand of  (\ref{prat}) is identified with the inverse metric. In addition, it can be seen that all the non linearities are polynomials in the variables $\eta_\alpha$ and $\partial_\mu \eta_\alpha$, with well behaved derivatives. From this, the following affirmation follows.

\begin{prop}\label{harmonicas}
Consider the Stelle's equations of motion 
$$
 \alpha\square G_{\mu\nu} +\frac{1}{16 \pi G_N}G_{\mu\nu} +(\alpha- 2\beta) RG_{\mu\nu}
+(2\beta-\alpha) (g_{\mu\nu} \square -\nabla_\mu\nabla_\nu) R
$$
$$
+2\alpha (R_{\mu\rho\nu\sigma} -\frac{1}{4} g_{\mu\nu}\, G_{\rho\sigma}) G^{\rho\sigma}+\frac{1}{2}(\alpha-2\beta) g_{\mu\nu}R^2=0,
$$
and the vector $\eta_\alpha=(F^\beta, \Phi_\gamma^\beta, \nabla_\mu \delta G^\mu_\nu)$ composed by the quantities (\ref{tearmo}), its derivatives $\Phi^\alpha_\beta=\partial_\alpha F_\beta$ and the covariant derivatives of the quantities $\delta G_{\mu\nu}$ defined in (\ref{gang}). Assume that a globally hyperbolic solution ($M_4$, $g_4$) of these equations  has been constructed, with an initial Cauchy hypersurface $\Sigma$.
Then, if initially $\eta_\alpha|_{t=0}=\partial_t \eta_\alpha|_{t=0}=0$ in a subset $\Omega$ of $\Sigma$, it follows that all the derivatives of $\eta_\alpha$ will be zero and the solution will be zero in $D(\Omega)$. In particular, as the quantities $\eta_\alpha$ contain $F^\alpha$ as entries, this ensures that $F^\alpha=0$ in $D(\Omega)$.
\end{prop}

It is therefore concluded from Proposition \ref{harmonicas} that if the initial constraints described above are satisfied in $\Omega$, then  $F^\alpha=0$ in $D(\Omega)$. Therefore, the use of harmonic coordinates is justified for the Stelle's gravity models. Nevertheless, as it will be discussed below, these constraints are too restrictive and some of them can be relaxed without spoiling the harmonic property.

\subsection{A closer look to the initial conditions for harmonic coordinates}
In the present section, the initial harmonic conditions of Proposition \ref{harmonicas} will be discussed in detail. From now, the discussion is focused on globally hyperbolic solutions  $(M, g)$. This means that the space-time $M$ is foliated by Cauchy surfaces $\Sigma_t$ determined in terms of a regular scalar function $t$, where the word "regular" means that its gradient is never vanishing.  Denote the initial Cauchy surface at $t=0$ by $\Sigma$. Then it follows from (\ref{peta}) that the initial conditions $\eta_\alpha|_{t=0}=\partial_t \eta_\alpha|_{t=0}=0$ at a subset $\Omega$ of $\Sigma$ are given by
$$
F^\alpha|_{t=0}=0,\qquad \dot{F}^\alpha|_{t=0}=0,
$$
\be\lb{const}
\nabla_\nu \delta G_\mu^\nu|_{t=0}=0,\qquad \dot{(\nabla_\nu \delta G_\mu^\nu)}|_{t=0}=0.
\ee
It is important to remark that the constraints $\Phi_\alpha^\beta|_{t=0}=0$ and $\dot{\Phi}_\alpha^\beta|_{t=0}=0$ for the quantity $\Phi_\alpha^\beta=\partial_\alpha F^\beta$ defined in (\ref{idol2}) have not been included. The reason  is that the spatial derivatives of $F^\alpha$ on the initial surface $\Sigma$ are all zero, thus $\Phi_\alpha^\beta|_{t=0}=0$ gives the same information as the second equation in (\ref{const}) and $\dot{\Phi}_\alpha^\beta|_{t=0}=0$ gives the same information as the third equation in (\ref{const}), therefore they can be safely omitted.  

The last three constraints in (\ref{const}) are not present in the GR. This reflects the fact that the Stelle's equations are of higher order. In order to clarify their meaning, consider the formula (\ref{id12}) for $\nabla_\nu \delta G_\mu^\nu$. As initially $\partial_i F^\alpha$ and $\partial_t F^\alpha=0$ the third constraint in (\ref{const}) implies that $\partial^2_t F^\alpha=0$ at $t=0$. The remaining constraint implies that $\partial^3_t F^\alpha=0$. Once these constraints are fulfilled, $F^\alpha=0$ during the evolution, which is the desired property for the use of the harmonic gauge. 

However, there is an odd feature in the aforementioned constraints. The quantities $F^\alpha$ contain first time derivatives of the metric $g_{\mu\nu}$ and therefore, the last condition $\partial^3_t F^\alpha=0$ may contain time derivatives of $g_{\mu\nu}$ up to fourth order. On the other hand, the equations of motion (\ref{gorila4}) are itself of fourth order, and this may indicate an inconsistency in the use of harmonic coordinates, unless this constraint is shown to be redundant. This problem would be solved if, given the initial conditions $F^\alpha=\partial_t F^\alpha=\partial^2_t F^\alpha=0$ at $\Sigma$, the condition $\partial^3_tF^\alpha=0$ comes out as a consequence of the equation of motion (\ref{gorila4}) in harmonic coordinates. 

Before proving this redundancy, let us recall that the Stelle's equations in the harmonic gauge are given by $H_{\mu\nu}^F=0$. These equations are supplemented with the constraints (\ref{gorilap4})-(\ref{gorilapp4}). From the definition $H_{\mu\nu}^F=H_{\mu\nu}-\delta H_{\mu\nu}$, it follows that the Stelle's equations in harmonic coordinates  are equivalent to \be\lb{istelo}H_{\mu\nu}=\delta H_{\mu\nu}.\ee The last expression is more practical, since the initial constraints were formulated above in terms of $H_{\mu\nu}$, not in terms of $H_{\mu\nu}^F$. In fact, the results of the previous section show that the initial constraints are $$n^\mu n^\nu H_{\mu\nu}|_{t=0}=0,\qquad n^\mu h^\nu_\alpha H_{\mu\nu}|_{t=0}=0.$$ From this, together with the fact that the equations of motion are equivalent to $H_{\mu\nu}=\delta H_{\mu\nu}$, it is easily found that the initial conditions can be cast in the following form
\be\lb{menso}
n^\mu n^\nu \delta H_{\mu\nu}|_{t=0}=0,\qquad n^\mu h^\nu_\alpha \delta H_{\mu\nu}|_{t=0}=0.
\ee
The advantage of the formulas (\ref{menso}) is that the quantity $\delta H_{\mu\nu}$ is a expression given in terms of $F^\alpha$, as shown in the definition (\ref{difer}). This fact will be helpful for showing that $\partial^3_t F^\alpha=0$ is redundant. In addition,
it is convenient to write down the explicit expression of $\delta H_{\mu\nu}|_{t=0}$.
Since it is assumed that initially $F^\alpha=0$ and $\partial_\mu F^\alpha=0$, it follows from (\ref{difer}) that
$$
\delta H_{\mu\nu}|_{t=0}=\alpha\square \delta G_{\mu\nu}|_{t=0}+(2\beta-\alpha)(g_{\mu\nu}\square-\nabla_\mu\nabla_\nu)(\partial_\alpha F^\alpha)|_{t=0}.
$$ 
On the other hand, from the formula
$$
\nabla_\nu\nabla_\nu \partial_\alpha F^\alpha=\partial_{\nu}\partial_\mu \partial_\alpha F^\alpha+\Gamma_{\mu\nu}^\beta \partial_\beta\partial_\alpha F^\alpha=\delta_{\mu t}\delta_{\nu t}\partial^3_t F^t
+\Gamma_{\mu\nu}^t \partial_t^2 F^\alpha,
$$
together with the harmonic condition $F^\alpha=\Gamma^\alpha=0$ at $t=0$ it is obtained that
$$
\square \partial_\alpha F^\alpha|_{t=0}=g^{\gamma\delta}\partial_{\gamma}\partial_\delta \partial_\alpha F^\alpha|_{t=0}=g^{00}\partial^3_t F^0|_{t=0}.
$$
Furthermore, it is not difficult to check that
$$
\nabla_\alpha\nabla_\beta G_{\mu\nu}=\nabla_\alpha [\partial_\beta G_{\mu\nu}+\Gamma^{\gamma}_{\beta \mu}G_{\gamma\nu} +\Gamma^{\gamma}_{\beta \nu}G_{\gamma\mu}+\Gamma^{\gamma}_{\nu \mu}G_{\gamma\beta}]=\partial_\alpha\partial_\beta G_{\mu\nu}+L_1(G_{\gamma\delta})+ L_2(\partial_\epsilon G_{\gamma\delta}),
$$
where $L_i$ are homogeneous and of first order in its arguments. Then, from the definition
$$
\delta G_{\mu\nu}=\frac{1}{2}(
g_{\mu\beta}\partial_\nu F^\beta+g_{\nu\beta}\partial_\mu F^\beta-g_{\mu\nu}\partial_\alpha F^\alpha),
$$
it is obtained that $\delta G_{\mu\nu}|_{t=0}=0$.

At this point, the initial condition $\partial^2_t F^\alpha|_{t=0}=0$ was ignored. Assume from now that this condition holds. The task is to show that $\partial^3_t F^\alpha|_{t=0}=0$ due to the whole set of initial constraints.
First of all, it should be noted that $\partial^2_t F^\alpha|_{t=0}=0$ implies that $\partial_t \delta G_{\mu\nu}|_{t=0}=0$. Furthermore
$$
\nabla_\nu\nabla_\nu \partial_\alpha F^\alpha|_{t=0}=\delta_{\mu 0}\delta_{\nu 0}\partial^3_t F^0|_{t=0}.
$$
Therefore, it follows that
$$
\delta H_{\mu\nu}|_{t=0}=\alpha\square \delta G_{\mu\nu}|_{t=0}+(2\beta-\alpha)(g_{\mu\nu}\square-\nabla_\mu\nabla_\nu)(\partial_\alpha F^\alpha)|_{t=0}=g^{00}\partial_t^2 \delta G_{\mu\nu}|_{t=0}
$$
$$
+(2\beta-\alpha)(g_{\mu\nu}g^{00}\partial^3_t -\delta_{\mu 0}\delta_{\nu 0}\partial^3_t)F^t|_{t=0}.
$$ 
After deriving this formula, the next step is to impose (\ref{menso}). But the quantities $F^\alpha$ are only time dependent in $\Sigma$. In this situation, the constraints (\ref{menso}) are satisfied if and only if  $\delta H_{0\mu}|_{t=0}=0$.
Suppose first the simplest situation namely, the one for which $g_{0i}=0$. Then, by taking into account the already assumed initial conditions, the three constraints $\delta H_{0i}|_{t=0}=0$ with $i=1,2,3$ are equivalent to
$$
g^{00}\partial_t^2 \delta G_{0i}|_{t=0}=\frac{1}{2}g^{00}\partial_t^2(g_{ij}\partial_t F^j-g_{0i}\partial_t F^0)|_{t=0}=\frac{1}{2}g^{00}g_{ij}\partial^3_t F^j|_{t=0}=0.
$$
These are three homogeneous equations and, if the determinant of the spatial metric $g_{ij}$ is non zero, then $\partial^3_t F^j=0$.
On the other hand, the constraint $\delta H_{0\mu}|_{t=0}=0$ implies that
$$
g^{00}\partial_t^2 \delta G_{00}|_{t=0}
+(2\beta-\alpha)(g_{00}g^{00}\partial^3_t -\partial^3_t)F^t|_{t=0}=g^{00}\partial_t^2 \delta G_{\mu\nu}|_{t=0}
$$
$$
=\frac{1}{2}g^{00}\partial_t^2(
2g_{0\beta}\partial_t F^\beta-g_{00}\partial_t F^0)|_{t=0}=-\frac{1}{2}g^{00}g_{00}\partial^3_t F^0|_{t=0}=0,
$$
where the initial conditions $F^\alpha=\partial_t F^\alpha=\partial_t^2 F^\alpha=0$ have been taken into account. The last formula shows that $\partial^3_t F^0=0$. Thus, the initial conditions $\partial^3_tF^\alpha|_{t=0}=0$ are direct  consequences of the Stelle's equation and can be safely ignored, when $g_{0i}|_{t=0}=0$.

Consider now the opposite situation namely, the one for which $g_{0i}|_{t=0}\neq 0$. In this case, the  component $\delta H_{00}|_{t=0}$ leads to
$$
g^{00}\partial_t^2 \delta G_{00}|_{t=0}+(2\beta-\alpha)(g_{00}g^{00}\partial^3_t -\partial^3_t)F^t|_{t=0}=\frac{1}{2}g^{00} (g_{00}\partial^3_t F^0-g_{0i}\partial^3_t F^i)|_{t=0}
$$
$$
+(2\beta-\alpha)(g_{00}g^{00}\partial^3_t -\partial^3_t)F^t|_{t=0}=0.
$$
On the other hand, the projection  $\delta H_{0i}|_{t=0}$ is explicitly
$$
g^{00}\partial_t^2 \delta G_{0i}|_{t=0}+(2\beta-\alpha)g_{0i}g^{00}\partial^3_t F^t|_{t=0}=\frac{1}{2}(
g_{i\beta}\partial^3_t F^\beta
-g_{0i}\partial^3_t F^t)|_{t=0}
$$
$$
+(2\beta-\alpha)g_{0i}g^{00}\partial^3_t F^t|_{t=0}=0.
$$
The last system of equations is homogeneous and, for a generic initial metric, it has a non zero determinant. Therefore $\partial_t^3 F^\alpha|_{t=0}=0$ when $g_{0i}\neq 0$.
   
   From the above discussion, it is concluded that the initial conditions for the harmonic gauge are 
\be\lb{const2}
F^\alpha|_{t=0}=0,\qquad \dot{F}^\alpha|_{t=0}=0,\qquad
\nabla_\nu \delta G_\mu^\nu|_{t=0}=0,
\ee
or equivalently $F^\alpha|_{t=0}=\partial_t F^\alpha|_{t=0}=\partial^2_t F^\alpha|_{t=0}=0$ for $\alpha=1,2,3,4$. The condition $\partial^3_t F^\alpha|_{t=0}=0$ or equivalently, $\dot{(\nabla_\nu \delta G_\mu^\nu)}|_{t=0}=0$, is therefore redundant. These results can be expressed shortly in the form of the following proposition.
  
 \begin{prop}\label{harmonicas2}
Suppose that a globally hyperbolic solution $(M_4,g_4)$ of the Stelle's equations (\ref{gorila4}) has been found, with
$\Sigma$ an initial Cauchy surface. If the conditions
$$
F^\alpha|_{t=0}=0,\qquad \dot{F}^\alpha|_{t=0}=0,\qquad
\nabla_\nu \delta G_\mu^\nu|_{t=0}=0,
$$
are satisfied in a subset $\Omega$ of $\Sigma$, with $\delta G_{\mu\nu}$ defined in (\ref{gang}), then $F^\alpha=0$ in the Cauchy development $D(\Omega)$.
\end{prop}

This proposition is essentially the same as Proposition \ref{harmonicas}, but with the redundant initial conditions removed.

\section{The degrees of freedom of the theory}
In the previous section,  the consistency of the use of harmonic coordinates when dealing with the Stelle's equations was pointed out. The next step is to understand the properties of the solutions of the model, when this gauge is imposed. The Stelle's equations  (\ref{gorila4}) are non linear of fourth order for the unknown metric $g_{\mu\nu}$. A possible approach for dealing with these equations is to convert them into a larger system of lower order. But before to employ this procedure, it may be convenient to identify the degrees of freedom of the Stelle's model, since they may give a hint about which variables should be taken as independent when making the order reduction procedure. The material of this section is not mandatory and the more mathematically oriented reader may skip to the next sections. However, the intuition beyond the mathematical formalism of these sections is inspired from the present considerations. The degrees of freedom in Stelle's gravity were already classified in  \cite{stelle}. For the present purposes however, a more suited reference to follow is  \cite{hinterblicher}.
In order to clarify these degrees, it is convenient to cast the action (\ref{bacho}) into the following equivalent form
\be\label{bach}
S=M_P^2\int d^4x\sqrt{-g}\left[\frac{1}{2}R+\frac{1}{12m^2}R^2+\frac{1}{4M^2}C_{\mu\nu\rho\sigma}C^{\mu\nu\rho\sigma}\right].
 \ee
 The equivalence between (\ref{bach}) and (\ref{bacho}) follows from the fact that the Gauss-Bonnet term $R_{\mu\nu\rho\sigma}R^{\mu\nu\rho\sigma}-4R_{\mu\nu}R^{\mu\nu}+R^2$ does not contribute to the equations of motion. The advantage of expressing the action in the form (\ref{bach}) is that the Weyl tensor 
 \be\lb{weyl}
 C_{\mu\nu\alpha\beta}=R_{\mu\nu\alpha\beta}+ R_{\mu[\alpha}g_{\nu]\beta}- R_{\beta[\alpha}g_{\nu]\mu}+\frac{1}{3}R g_{\mu[\alpha}g_{\beta]\nu},
 \ee
 is a conformal invariant. This property can be exploited to understand the degrees of freedom of the model as follows. Note first that the last action is equivalent to the following one
\be\label{bach2}
S={M_P^2}\int d^4x \sqrt{-g}\left[\frac{1}{2}\left(1+{\phi\over 3m^2}\right)R-\frac{1}{12m^2}\phi^2+\frac{1}{4M^2}C_{\mu\nu\rho\sigma}C^{\mu\nu\rho\sigma}\right].
 \ee
The equivalence follows from the fact that the $\phi$ equation of motion gives that $\phi=R$, which upon substitution into (\ref{bach2}) returns (\ref{bach}).
A transformation of the metric of the form $g_{\mu\nu}\to \Omega^2 g_{\mu\nu}$ has no effect on the Weyl invariant  term $C_{\mu\nu\rho\sigma}C^{\mu\nu\rho\sigma}$. The specific choice 
$$ g_{\mu\nu}\rightarrow \frac{3m^2}{\phi+3m^2}g_{\mu\nu},$$
followed by a field redefinition 
\be\lb{rede}\phi=3m^2\left(e^\psi-1\right),
\ee
gives that $g_{\mu\nu}\rightarrow e^{-\psi}g_{\mu\nu}$. The action in the new frame becomes
$$ 
S=M_P^2\int d^4x \sqrt{-g}\left[\frac{1}{2}R-\frac{3}{4}(\partial\psi)^2-\frac{3}{4}m^2e^{-2\psi}\left(e^\psi-1\right)^2+\frac{1}{4M^2}C_{\mu\nu\rho\sigma}C^{\mu\nu\rho\sigma}\right].
$$
Next, in order to eliminate the Weyl term squared part $C_{\mu\nu\rho\sigma}C^{\mu\nu\rho\sigma}$, it may be helpful to introduce a symmetric dimensionless auxiliary tensor field $f_{\mu\nu}$. The action is then transformed into
\be\lb{akshon}
 S={M_P^2}\int d^4x \sqrt{-g}\left[\frac{1}{2}R-\frac{3}{4}(\partial\psi)^2-\frac{3}{4}m^2e^{-2\psi}\left(e^\psi-1\right)^2 +f^{\mu\nu}G_{\mu\nu}-\frac{1}{2}M^2\left(f_{\mu\nu}f^{\mu\nu}-f_\rho^\rho f_\eta^\eta\right)\right],
\ee
 where $G_{\mu\nu}$ is the Einstein tensor of $g_{\mu\nu}$, and indices are always lowered with the help of $g_{\mu\nu}$.
The equations of motion for $f_{\mu\nu}$ give \be\lb{fanta}f_{\mu\nu}=\frac{1}{M^2}\left(R_{\mu\nu}-\frac{1}{6}Rg_{\mu\nu}\right).
\ee When this is inserted into (\ref{akshon}) the original action is obtained, up to a term proportional to the Gauss-Bonnet invariant $R_{\mu\nu\rho\sigma}R^{\mu\nu\rho\sigma}-4R_{\mu\nu}R^{\mu\nu}+R^2$ which does not modify the equations of motion.

      The formulation in (\ref{akshon}) is manifestly of second order. A hint to deal with the Cauchy problem for this model comes from the analysis of the degrees of freedom. These degrees are elucidated by  a second order expansion around the Minkowski background $g_{\mu\nu}=\eta_{\mu\nu}$, $f_{\mu\nu}=0$, $\psi=0$ of the action (\ref{akshon}), up to second order terms. The perturbation is expressed as \be\lb{pertu}g_{\mu\nu}=\eta_{\mu\nu}+h_{\mu\nu},\qquad f_{\mu\nu}=\Psi_{\mu\nu},\qquad \psi=\chi,\ee where the quantities $h_{\mu\nu}$, $\Psi_{\mu\nu}$ and $\chi$ represent small deviations from the trivial vacuum.  The expanded action is given by
$$ S_2={M_P^2}\int d^4x\bigg[-{3\over 4}\left((\partial\chi)^2+m^2\chi^2\right) +{1\over 8}h^{\mu\nu}\left({\mathcal E}h\right)_{\mu\nu}-{1\over 2}\Psi^{\mu\nu}\left({\mathcal E}h\right)_{\mu\nu}-{1\over 2}M^2\left(\Psi_{\mu\nu}\Psi^{\mu\nu}-\Psi_\rho^\rho \Psi_\eta^\eta\right)\bigg],
$$
where $$\left({\mathcal E}h\right)_{\mu\nu}\equiv\square h_{\mu\nu}-\eta_{\mu\nu}\square h-2\partial_{(\mu}\partial^\rho h_{\nu)\rho}+\partial_\mu\partial_\nu h+\eta_{\mu\nu}\partial^\rho\partial^\sigma h_{\rho\sigma},$$ is the  graviton kinetic operator. The analogous operator was introduced for the perturbation $\Psi_{\mu\nu}$. The tensor kinetic terms in the last action can be broken to a diagonal form by the use of the field redefinition 
$$
h_{\mu\nu}=2\left(h_{\mu\nu}'+\Psi_{\mu\nu}\right),
$$
for which the action is converted into
$$ 
S=\int d^4x \bigg[-\frac{3}{4}\left((\partial\chi)^2-m^2\chi^2\right) +\frac{1}{2}h'^{\mu\nu}\left({\mathcal E} h'\right)_{\mu\nu}-\frac{1}{2}\Psi^{\mu\nu}\left({\mathcal E} \Psi\right)_{\mu\nu}-\frac{1}{2}M^2\left(\Psi_{\mu\nu}\Psi^{\mu\nu}-\Psi_\rho^{\;\;\rho} \Psi_\eta^{\;\;\eta}\right)\bigg].
$$ 
This linearized action was found in \cite{stelle2}. It shows that the degrees of freedom are composed by a scalar field $\chi$, a massless spin two field $h'_{\mu\nu}$ and a spin two massive field $\Psi_{\mu\nu}$ whose kinetic energy has opposite sign to that of the graviton. The signs of the kinetic terms of the spin two fields may be interchanged, but it is likely that instabilities will not be avoided by this procedure.

\section{The Stelle's equations as a second order quasi-linear hyperbolic system}

\subsection{A reduction order procedure for the Stelle's equations} 

The previous discussion shows that the spin two degree of freedom $\Psi_{\mu\nu}$ is identified with $$f_{\mu\nu}=R_{\mu\nu}-\frac{1}{6}g_{\mu\nu}R,$$
when the space-time metric $g_{\mu\nu}=\eta_{\mu\nu}+h_{\mu\nu}$ is close enough to a Minkowksi one $\eta_{\mu\nu}.$
In the same approximation, the identification $\phi=R$  in the previous section, together with the formula (\ref{rede}), show that scalar degree of freedom $\psi=\chi$ is proportional to the Ricci scalar curvature $R$ for small values. Furthermore, the massless spin two field $h'_{\mu\nu}$ is identified with the graviton. This suggests that it may be advantageous to find a system of equations of lower order, completely equivalent to (\ref{ecmov}), in which the unknowns are constituted by the curvature $f_{\mu\nu}$, the scalar curvature $R$ and the metric $g_{\mu\nu}$. This may reduce the order of the system, thus making it more tractable. It will be more practical however, to take the traceless part  
$$
\widetilde{R}_{\mu\nu}=R_{\mu\nu}-\frac{1}{4}g_{\mu\nu}R,
$$
of the Ricci curvature $R_{\mu\nu}$, the metric $g_{\mu\nu}$ and the scalar curvature $R$ as the independent unknowns \cite{noakes}. 

A system with these properties is derived below, and differs slightly with the one found in \cite{noakes}.
However, our conclusions agree with that reference. The advantage of the present procedure is to make explicit some terms the author \cite{noakes} does not specify. This explicit form is of importance for applying modern theorems, and for making statements about the regularity of the solutions of the model. In addition, these terms may be relevant for studying stability problems.

Before obtaining the desired system, it is convenient to express the equations of motion (\ref{gorila4}) in the following form
$$
\frac{1}{16\pi G_N} (R_{\mu\nu} -\frac{1}{2} R g_{\mu\nu}) - 2\beta R(R_{\mu\nu}-\frac{1}{4}R g_{\mu\nu})
+(2\beta-\alpha) (g_{\mu\nu} \square -\nabla_\mu\nabla_\nu) R
$$
\be\lb{gorila}
+ \alpha\square(R_{\mu\nu}-\frac{1}{2} R \, g_{\mu\nu})
+
2\alpha (R_{\mu\rho\nu\sigma} -\frac{1}{4} g_{\mu\nu}\, R_{\rho\sigma}) R^{\rho\sigma} =0.
\ee
The trace of (\ref{ecmov}) contains some terms proportional to $R^2$ and to $R_{\mu\nu}R^{\mu\nu}$, but these terms cancel each other out. 
The resulting equation for $R$ is simply
\begin{equation}\label{descom3}
2(3\beta-2\alpha)\square R-\frac{1}{16\pi G_N}R=0.
\end{equation}
The equation for the traceless component $\widetilde{R}_{\mu\nu}$ can be found by multiplying the equation (\ref{descom3}) by $g_{\mu\nu}/4$ and subtracting it from (\ref{gorila}), yielding the result
$$
\frac{1}{16\pi G_N} (R_{\mu\nu} -\frac{1}{4} R g_{\mu\nu}) - 2\beta R(R_{\mu\nu}-\frac{1}{4}R g_{\mu\nu})
+(2\beta-\alpha) (\frac{1}{4}g_{\mu\nu} \square -\nabla_\mu\nabla_\nu) R
$$
\be\lb{gorilaz}
+ \alpha\square(R_{\mu\nu}-\frac{1}{4} R \, g_{\mu\nu})
+
2\alpha (R_{\mu\rho\nu\sigma} -\frac{1}{4} g_{\mu\nu}\, R_{\rho\sigma}) R^{\rho\sigma} =0.
\ee
Alternatively, the last system can be written as follows
$$
(2\beta-\alpha) (\frac{1}{4}g_{\mu\nu} \square -\nabla_\mu\nabla_\nu) R+ \alpha\square\widetilde{R}_{\mu\nu}+\frac{1}{16\pi G_N} \widetilde{R}_{\mu\nu} + (\alpha-2\beta) R\widetilde{R}_{\mu\nu}
$$
\be\lb{gorilaz0}
+
2\alpha (R_{\mu\rho\nu\sigma} -\frac{1}{4} g_{\mu\nu}\, \widetilde{R}_{\rho\sigma}) \widetilde{R}^{\rho\sigma} =0.
\ee
By taking (\ref{descom3})
into account, the last expression becomes
$$
 \bigg(\alpha\square+\frac{1}{16\pi G_N}\bigg)\widetilde{R}_{\mu\nu}+(2\beta-\alpha) \bigg[\frac{1}{128\pi G_N(3\beta-2\alpha)}g_{\mu\nu} -\nabla_\mu\nabla_\nu\bigg] R + (\alpha-2\beta) R\widetilde{R}_{\mu\nu}
$$
\be\lb{descom4}
+
2\alpha (R_{\mu\rho\nu\sigma} -\frac{1}{4} g_{\mu\nu}\, \widetilde{R}_{\rho\sigma}) \widetilde{R}^{\rho\sigma}=0.
\ee
Note that the trace of (\ref{descom4}) is zero, as it should be.

In equations (\ref{descom3})-(\ref{descom4})  the quantities $g_{\mu\nu}$, $\widetilde{R}_{\mu\nu}$ and $R$ are not independent unknowns, which is the desired feature. Thus, further work has to be done. Some remarks are in order. The system to be constructed below is equivalent to  
(\ref{descom3})-(\ref{descom4}) when harmonic coordinates are employed. However, the use of the harmonic is legitimate when dealing with equations (\ref{gorila4}) or equivalently (\ref{descom3})-(\ref{descom4}), this was justified in the Propositions 1 and 2 of the previous sections. The point of using this gauge is that it simplifies considerably the resulting equations.   

The first equation to be introduced is simply
\begin{equation}\label{desco}
-\frac{1}{2}g^{\eta\delta}g_{\mu\nu,\eta\delta}+Q_{\mu\nu}(g,\partial g)=\widetilde{R}^F_{\mu\nu}
+\frac{1}{4}g_{\mu\nu}R^F,
\end{equation}
which is valid for harmonic coordinates. The meaning of this identity is transparent. The left hand side is the expression of the Ricci tensor $R_{\mu\nu}^F$ in terms of the metric $g_{\mu\nu}$ in harmonic coordinates (\ref{armonia})-(\ref{armonia2}), and the right hand is the same quantity expressed in terms of $\widetilde{R}^F_{\mu\nu}$ and $R^F$.
This is a second order equation for $g_{\mu\nu}$, in which the quantities $R^F$ and $\widetilde{R}^F_{\mu\nu}$ in the right hand side are acting as sources.

The remaining equations are obtained as follows. Replace in (\ref{gorila}) the quantities $2\beta R$ and  $4R_{\mu\nu\alpha\beta}-g_{\mu\nu}R_{\rho\sigma}$ in terms of the metric $g_{\mu\nu}$, by assuming harmonic coordinates. Replace the other curvature expressions by its $R^F$ or $R_{\mu\nu}^F$ counterparts.  Then consider the traceless part and the trace part of the resulting equation by taking into account that $R^F=g^{\mu\nu}R^F_{\mu\nu}$.
The result are the two following equations
\begin{equation}\label{desco4}
2(3\beta-2\alpha)\square R^F-\frac{1}{16\pi G_N} R^F-\alpha (g_{\mu\rho}\partial_\sigma F^\mu+g_{\mu\sigma}\partial_\rho F^\mu)(\widetilde{R}^{F\rho\sigma}+\frac{1}{4}g^{\rho\sigma}R^F)=0,
\end{equation}
$$
\alpha \square \widetilde{R}^F_{\mu\nu}-(2\beta-\alpha) \nabla_{\mu}\nabla_{\nu}R^F+\frac{1}{16\pi G_N}\widetilde{R}^F_{\mu\nu}+ 
\frac{(2\beta-\alpha)}{128\pi G_N(3\beta-2\alpha)}g_{\mu\nu} R^F 
$$
$$
+ (\alpha-2\beta)\bigg[-g^{\alpha\beta}g^{\sigma\rho}g_{\sigma\rho,\alpha\beta}+2Q(g,\partial g)\bigg]\widetilde{R}^F_{\mu\nu}
+\frac{\alpha}{4} g_{\mu\nu}\bigg[-g^{\alpha\beta}g_{\rho\sigma,\alpha\beta}+2Q_{\rho\sigma}(g,\partial g)\bigg]g^{\eta\rho}g^{\delta\sigma} \widetilde{R}^F_{\eta\delta}
$$
\be\lb{desco3}
+
\alpha \bigg[g_{\rho\sigma,\mu\nu} +g_{\mu\nu,\rho\sigma}-g_{\rho\nu,\mu\sigma} -g_{\mu\sigma,\rho\nu}+2g_{\alpha\beta}(\Gamma^{\alpha}_{\rho\nu}\Gamma^{\beta}_{\mu\sigma}-\Gamma_{\rho\sigma}^{\alpha}\Gamma_{\mu\nu}^{\beta})\bigg]g^{\eta\sigma}g^{\delta\rho}\widetilde{R}^F_{\eta\delta}=0.
\ee 
The terms proportional to the derivatives $\partial_\alpha F^\beta$ in (\ref{desco4}) arise due to the fact that $4R_{\mu\sigma\nu\rho}-g_{\mu\nu}R_{\rho\sigma}$ is traceless with respect to the indices $\mu$ and $\nu$ for any coordinate system but instead, $4R_{\mu\rho\nu\sigma}-g_{\mu\nu}R^F_{\rho\sigma}$ is not. 
The expression for the trace follows from (\ref{desco2}) and is the one inducing the last term in (\ref{desco4}). However, these terms may be neglected if the harmonic condition $F^\alpha=0$ is employed.

The unknowns for the system (\ref{desco})-(\ref{desco3}) are $g_{\mu\nu}$,  $\widetilde{R}^F_{\mu\nu}$ and  $R^F$, and these equations are equivalent to  (\ref{descom3})-(\ref{descom4}). These facts are collected in the following proposition.

\begin{prop}\label{equivalencia}
Consider a triple composed by a $C^4$ metric $g_{\mu\nu}$, a  $C^2$ tensor $\widetilde{R}^F_{\mu\nu}$ and a $C^2$ scalar $R^F$, all these quantities defined in a 4-dimensional domain $\Omega$ of a given space-time $M_4$. The metric  $g_{\mu\nu}$ is such that the harmonic condition $F^\alpha=0$ is satisfied and furthermore $|g_{\mu\nu}|\neq 0$ in $\Omega$.
If this triple is a local solution of the system (\ref{desco})-(\ref{desco3}) then the metric $g_{\mu\nu}$ is also a solution of the system (\ref{descom3})-(\ref{descom4}), when the curvatures $\widetilde{R}_{\mu\nu}$ and $R$ are expressed in terms of $g_{\mu\nu}$ by the standard formulas of differential geometry. Conversely, given a $C^4$ metric $g_{\mu\nu}$ which satisfies the harmonic condition $F^\alpha=0$ and (\ref{descom3})-(\ref{descom4}) in the domain $\Omega$, consider the quantities $\widetilde{R}^F_{\mu\nu}$ and $R^F$ constructed in terms of the standard formulas of differential geometry. These tensors are $C^2$ and the triple $g_{\mu\nu}$, $\widetilde{R}^F_{\mu\nu}$ and $R^F$ solve the system (\ref{desco})-(\ref{desco3}) in $\Omega$.  The metric $g_{\mu\nu}$ is also a solution of the Stelle's classical equations of motion (\ref{gorila4}) in $\Omega$.
\end{prop}

\emph{Proof:} In order to prove this statement, find the trace and the traceless part of (\ref{desco}). This procedure gives the expressions of $\widetilde{R}^F_{\mu\nu}$ and $R^F$  in terms of the metric $g_{\mu\nu}$. Replace the result in (\ref{desco4})-(\ref{desco3}), then  the resulting equations are of fourth order in $g_{\mu\nu}$. On the other hand, replace all the curvatures in (\ref{descom3})-(\ref{descom4}) in terms of $g_{\mu\nu}$ by use of the standard expressions of differential geometry, by assuming the harmonic gauge. The resulting system is also of fourth order for $g_{\mu\nu}$ and, furthermore, it coincides with the previous one. In fact, the system (\ref{desco})-(\ref{desco3}) was constructed intentionally for this to happen. As the system (\ref{descom3})-(\ref{descom4}) is the trace and the traceless part the Stelle's equation of motions (\ref{gorila4}) in harmonic coordinates it follows, by taking into account the regularity of the metric and curvatures specified above, that the solutions of any of these systems will be solutions of the Stelle's gravity equations. $\square$
\\

At this point, it is perhaps worthy to discuss the subtle differences between the present approach and the one employed in \cite{noakes}. In \cite{noakes} the author presents a system corresponding to the three unknowns $g_{\mu\nu}$, $\widetilde{R}_{\mu\nu}$ and $R$. But it is just after this system is presented that the use of the harmonic coordinates is justified. For this reason, this author is forced to introduce calligraphic curvatures variables $\widetilde{\cal R}_{\mu\nu}$ and ${\cal R}$, and to prove after some work they are equal to the standard ones. In addition, it is not clear that the zero trace condition $g^{\mu\nu}\widetilde{\cal R}_{\mu\nu}=0$ is satisfied for the calligraphic curvatures, until they are shown to be equal to the standard ones. This complicates the analysis considerably. In the present approach instead, the use of harmonic coordinates is justified from the results of the previous sections and furthermore, the zero trace condition is ensured from the very beginning in (\ref{desco}). This is at the cost of the appearance of a  term proportional to the derivatives $\partial_\alpha F^\beta$ in (\ref{desco4}), which is absent in the system derived in the reference \cite{noakes}, even before justifying the use of harmonic coordinates. However, our conclusions about the Cauchy problem agree with that reference.  

Having agreed upon the conclusions of \cite{noakes}, our purpose is to give an alternative point of view and to enlarge the results given in this reference. In the present approach, all the resulting terms in the second order formulation are given explicitly. This permits the application of certain mathematical statements about the regularity of the solutions. In particular, this leads to the conclusion that, given $C^\infty$ initial conditions, there exists two time values $T_1<0$ and $T_2>0$ for which the universe evolution is $C^\infty$ in $(T_1,T_2)$. Another consequence is the  existence of a maximally hyperbolic development. The obtained system may be useful, in addition, for proving the stability of the solution under small perturbations of the initial conditions, although we have not a concrete proof of this fact.

\subsection{The hyperbolic second order quasi-linear form of the Stelle's system}

The system of equations (\ref{desco})-(\ref{desco3}) obtained above is  equivalent to the following one
\begin{equation}\label{descot}
-\frac{1}{2}g^{\eta\delta}g_{\mu\nu,\eta\delta}+Q_{\mu\nu}(g,\partial g)=\widetilde{R}_{\mu\nu}
+\frac{1}{4}g_{\mu\nu}R,
\end{equation}
\begin{equation}\label{descot4}
\square R-\frac{1}{32\pi G_N(3\beta-2\alpha)}R=0,
\end{equation}
$$
\alpha \square \widetilde{R}_{\mu\nu}-(2\beta-\alpha) \nabla_{\mu}\nabla_{\nu}R+\frac{1}{16\pi G_N}\widetilde{R}_{\mu\nu}+ 
\frac{(2\beta-\alpha)}{128\pi G_N(3\beta-2\alpha)}g_{\mu\nu}R  
$$
$$
+ (\alpha-2\beta)\bigg[-g^{\alpha\beta}g^{\sigma\rho}g_{\sigma\rho,\alpha\beta}+2Q(g,\partial g)\bigg]\widetilde{R}_{\mu\nu}
+\frac{\alpha}{4} g_{\mu\nu}\bigg[-g^{\alpha\beta}g_{\rho\sigma,\alpha\beta}+2Q_{\rho\sigma}(g,\partial g)\bigg] \widetilde{R}^{\rho\sigma}
$$
\be\lb{descot3}
+
\alpha \bigg[g_{\rho\sigma,\mu\nu} +g_{\mu\nu,\rho\sigma}-g_{\rho\nu,\mu\sigma} -g_{\mu\sigma,\rho\nu}+2g_{\alpha\beta}(\Gamma^{\alpha}_{\rho\nu}\Gamma^{\beta}_{\mu\sigma}-\Gamma_{\rho\sigma}^{\alpha}\Gamma_{\mu\nu}^{\beta})\bigg]\widetilde{R}^{\rho\sigma}=0,
\ee
after imposing term by term the harmonic condition $F^\alpha=0$.
In order to study formal properties of the system (\ref{descot})-(\ref{descot4}) it is convenient to convert it into a hyperbolic quasi-linear system of the form (\ref{choquet}). The advantage is that these systems are well studied in the literature. However,  the presence of terms such as $\nabla_\mu\nabla_\nu R$ or $g^{\alpha\beta}g_{\rho\sigma,\alpha\beta}$ in (\ref{descot3}) imply that the system is not quasi-linear. In other words, it is not of the form specified in (\ref{choquet}).  But this problem can be sorted out by introducing new variables
$r_\mu=\partial_\mu R$ and $c_{\mu\nu\alpha}=g_{\mu\nu,\alpha}$, and by taking derivatives of the equations (\ref{descot})-(\ref{descot4}) and adding them to the system. The resulting equations are
\begin{equation}\label{descoteno}
-\frac{1}{2}g^{\eta\delta}g_{\mu\nu,\eta\delta}=-Q_{\mu\nu}(g,\partial g)+\widetilde{R}_{\mu\nu}
+\frac{1}{4}g_{\mu\nu}R,
\end{equation}
\begin{equation}\label{descoteno4}
g^{\alpha\beta}\partial_{\alpha}\partial_{\beta}R=\frac{1}{32\pi G_N(3\beta-2\alpha)}R,
\end{equation}
\begin{equation}\label{descotem}
-\frac{1}{2}g^{\eta\delta}c_{\mu\nu\gamma,\eta\delta}=\frac{1}{2}g^{\eta\delta}_{\,\,\,\,,\gamma}\,c_{\mu\nu\eta,\delta}-Q_{\mu\nu,\gamma}(g,\partial g, c)+\widetilde{R}_{\mu\nu,\gamma}
+\frac{1}{4}g_{\mu\nu,\gamma}R+\frac{1}{4}g_{\mu\nu}r_\gamma,
\end{equation}
\begin{equation}\label{descotem4}
g^{\alpha\beta}\partial_{\alpha}\partial_{\beta}r_\gamma=\frac{1}{32\pi G_N(3\beta-2\alpha)}r_\gamma,
\end{equation}
$$
\alpha \square \widetilde R_{\mu\nu}=(2\beta-\alpha) \nabla_{\mu}r_\nu-\frac{1}{16\pi G_N}\widetilde{R}_{\mu\nu}-
\frac{(2\beta-\alpha)}{128\pi G_N(3\beta-2\alpha)}g_{\mu\nu}R 
$$
$$
 -(\alpha-2\beta)\bigg[-g^{\alpha\beta}g^{\sigma\rho}c_{\sigma\rho\alpha,\beta}+2Q(g,\partial g)\bigg]\widetilde{R}_{\mu\nu}
-\frac{\alpha}{8} g_{\mu\nu}\bigg[-g^{\alpha\beta}c_{\rho\sigma\alpha,\beta}+2Q_{\rho\sigma}(g,\partial g)\bigg]\widetilde{R}^{\rho\sigma}
$$
\be\lb{descoten3}
-
\alpha \bigg[c_{\rho\sigma\mu,\nu} +c_{\mu\nu\rho,\sigma}-c_{\rho\nu\mu,\sigma} -c_{\mu\sigma\rho,\nu}+2g_{\alpha\beta}(\Gamma^{\alpha}_{\rho\nu}\Gamma^{\beta}_{\mu\sigma}-\Gamma_{\rho\sigma}^{\alpha}\Gamma_{\mu\nu}^{\beta})\bigg]\widetilde{R}^{\rho\sigma}.
\ee
The fact that $F^\alpha=0$ was taken into account when writing these equations. 
\\

\emph{Lemma 1: The Stelle's system (\ref{descoteno})-(\ref{descoten3}) is of quasi-linear hyperbolic type, that is, it can be expressed as
\begin{equation}\label{chic}
g^{\mu\nu}(x,t,u)\partial_\mu\partial_\nu u_q(x,t)=f_q(x,t,u_i, \partial_\mu u_j),
\end{equation}
where $u_q$ with $q=1,..,n$ is a vector constituted by the $n$-unknowns and the matrix $g^{pq}$ is the same for all the equations $q=1,..n$ and it is of normal hyperbolic type, that is
$g^{44}\leq  0$ and $g^{ij}x_i x_j$ is a positive definite form for every point $x$ and $t$, with the latin indices indicating spatial directions. }
\\

\emph{Proof of Lemma 1:} In order to understand this statement, consider the vector $u_\alpha$ constituted by all the unknowns
\be\lb{viktor}
u_\alpha=(g_{\mu\nu}, R, r_\gamma, c_{\mu\nu\alpha}, \widetilde{R}_{\mu\nu}).
\ee
The laplacian
$\square \widetilde{R}_{\mu\nu}$ can be expressed as
$$
\square \widetilde{R}_{\mu\nu}=g^{\alpha\beta}\partial_\alpha\partial_\beta \widetilde{R}_{\mu\nu}+H_{\mu\nu}(\widetilde{R}_{\mu\nu}, \partial_\alpha \widetilde{R}_{\mu\nu}, \Gamma^\alpha_{\beta\gamma}),
$$
where the term $H_{\mu\nu}(\widetilde{R}_{\mu\nu}, \partial_\alpha \widetilde{R}_{\mu\nu}, \Gamma^\alpha_{\beta\gamma})$ is a linear combination of its arguments. The Christofell symbols $\Gamma^\alpha_{\beta\gamma}$ depend on the inverse metric $g^{\mu\nu}(g_{\alpha\beta})$ and its first derivatives $\partial_\alpha g_{\beta\gamma}$. By moving the term $H_{\mu\nu}$ to the right hand side in (\ref{descoten3}), it follows that the left hand side of the system (\ref{descoteno})-(\ref{descoten3}) is of the form $g^{\alpha\beta}\partial_\alpha\partial_\beta u_q$, which $g^{\mu\nu}$ the same everywhere, and equal to the inverse space-time metric. Thus $g^{44}\leq 0$ and $g^{ij}x_i x_j\geq 0$ with the latin indices denoting spatial directions. Therefore the system (\ref{descoteno})-(\ref{descoten3}) can be written in the form
$$
g^{\mu\nu}(x,t,u)\partial_\mu\partial_\nu u_q(x,t)=f_q(x,t,u_i, \partial_\mu u_j).
$$
This is of the form anticipated in the introduction (\ref{choquet}). The non linearity $f_q(x,t,u_i, \partial_\mu u_j)$ is a generic notation for the terms (\ref{descoteno})-(\ref{descoten3}) that do not contain second order derivatives. $\square$   
\\

In the next subsection, some general theorems about systems of the form (\ref{chic})  will be discussed. These results will be applied  to the Stelle gravity model, which is the main purpose of the present work.

\subsection{General results about second order hyperbolic quasi-linear systems}
 
In order to formulate some statements related to quasi-linear hyperbolic systems, the following relevant definitions are needed. 

Consider the map $g_{\mu\nu}\in C^{\infty}(\R^{nN+2N+n+1},L_n)$, where $L_n$ denotes the space of canonical
$(n+1)\times (n+1)$ Lorentz matrices. Assume that these quantities satisfy
\be\lb{cinf}
|\partial^\alpha g_{\mu\nu}(x,t, \xi)|\leq h_{I, \alpha}(|\xi|),
\ee
where ($x$ ,$t$) are local coordinates on $\R^{n+1}$ and $\xi$ parametrize the coordinates of $\R^{nN+2N}$. Here $I=[T_1, T_2]$ is any compact time interval and $h_{I, \alpha}: \R\to \R$ are continuous increasing functions for every  multi index $\alpha=$($\alpha_1$,..,$\alpha_{nN+2N+n+1})$. Suppose that, for any compact interval $I$, there are constants $a_i\geq 0$ with $i=1,2,3$ such that
\be\lb{cinf2}
g_{00}\leq -a_1,\qquad \det g_{ij}\geq a_2, \qquad \sum^n_{(\mu,\nu)=0}|g_{\mu\nu}|\leq a_3.
\ee
\emph{Definition: The quantities $g_{\mu\nu}$ satisfying the last condition are known as $C_{n, a}$ metrics, and the ones satisfying all of the aforementioned assumptions are known as $C^\infty$ $N$, $n$ admissible metrics.} 
\\

The second definition applies for the non linearity $f_q(x,t,u_i, \partial_\mu u_j)$. Assume that there exist some functions $\bar{h}_{I, \alpha}(|\xi|)$ such that the derivatives of the non linearity satisfy inequalities of the form 
\begin{equation}\label{chicas}
|\partial^\alpha f(x,t, \xi)|\leq \bar{h}_{I, \alpha}(|\xi|),
\end{equation}
with $\bar{h}_{I, \alpha}(|\xi|)$ functions of the same type as the  $h_{I, \alpha}(|\xi|)$ above. The time interval $I$ is supposed to be compact. In addition $f(x, t, \xi)$ is such that for each compact interval $I$,
there exist a compact set $K\subset \mathbb \R^3$ such that $f(x, t, 0)=0$ for any $x$ outside $K$ and $t \in I$. 
\\

\emph{Definition: Non linearities $f(x,t, \xi)$ which fulfil the conditions described above are known of locally of $x$-compact support.}
\\

The following proposition is of importance for studying the space of solutions of the Stelle's gravity model  \cite[Ch. 9]{ringstrom}.

\begin{prop}\label{stelle} Consider the hyperbolic quasi-linear second order system
$$
g^{\mu\nu}(x,t,u)\partial_\mu\partial_\nu u_q(x,t)=f_q(x,t,u_i, \partial_\mu u_j),
$$
with the initial conditions
\be\lb{chic 2}
u(x,T_0)=U_0,\qquad \partial_t u(x, T_0)=U_1.
\ee
Let $U_0, U_1 \in C^{\infty}(\R^n, \R^N)$  and $T_0\in\R$. Suppose that the quantity $g^{\mu\nu}(x,t,u)$ is a $C^\infty$ $N$, $n$ admissible metric, a concept that has been described in (\ref{cinf})-(\ref{cinf2}) and below. Furthermore, suppose that the quantity $f_q(x,t,u_i, \partial_\mu u_j)$ is locally of x-compact support, as described in (\ref{chicas}).  Then there exist two times $T_1$ and $T_2$ such that $T_1< T_0<T_2$ for which there exists a unique $C^\infty$ solution $u$ of the system \eqref{chic} and \eqref{chic 2}. This solution is of  $x$-compact support.
\end{prop}

It should be emphasized that the $x$-compact support is a rather technical one. Its importance resides in that a smooth function $u:R^{n+1}\to R^m$ with such property can be viewed as an element in $C^l[R, H^k(n, m)]$ for every value of $l$ and $k$. This plays an important role in the proof of the proposition, as shown in chapters 8 and 9 of \cite{ringstrom}.

\subsection{The role of x-compactness in the Stelle's system}

The Proposition \ref{stelle} given above ensures the existence of $C^\infty$ solutions for the equations (\ref{choquet}) when the non-linearity is of x-compact support, if suitable initial conditions are imposed. This is an important result, and it is worthy  to investigate whether the non  linearity $f_q(x,t, u_i, \partial_\mu u_j)$ defined by the equations (\ref{descoteno})-(\ref{descoten3}) is of x-compact support. This non linearity corresponds to all the terms in these equations which are not of second order. To give an example, from (\ref{descoteno}) it follows that
$$
f_1(u_\mu, \partial_\nu u_\mu)=Q_{\mu\nu}(g,\partial g)-\widetilde{R}_{\mu\nu}
-\frac{1}{4}g_{\mu\nu}R.
$$
The other components $f_q(u_\mu, \partial_\nu u_\mu)$ are similarly found from (\ref{descoteno4})-(\ref{descoten3}).

\begin{prop}\label{xcomp} By a metric redefinition $g^{\mu\nu}\to \overline{g}^{\mu\nu}= |g_{\mu\nu}|^n g_{\mu\nu}$ in (\ref{descoteno})-(\ref{descoten3}), with $n$ an appropriate integer, the system (\ref{descoteno4})-(\ref{descoten3}) is converted into one for which the resulting non linearity $\widetilde{f}_q(u_\mu, \partial_\nu u_\mu)$ is locally of x-compact support.
\end{prop}

 The following elementary property about polynomials in several variables is needed in order to prove this result.
 \\
 
\emph{Lemma 2: For any  polynomial $P(\xi_a)$ in several variables $\xi^a$ with $a=1,..., m$, there exists a single variable polynomial $Q(x)$ with positive coefficients such that $|P(\xi_a)|\leq Q(|\xi|)$, with $|\xi|$ the usual Euclidean norm of the vector $\xi=(\xi^1,.., \xi^m)$.}
\\

\emph{Proof of the Lemma 2:} Any polynomial in several variables $\xi^a$ with $a=1,..., m$ can be written in generic form as
$$
P(\xi^a)=\sum_{q_1,..,q_m=0}^{N_1,..,N_m} a_{q_1..q_m}(\xi^1)^{q_1} (\xi^2)^{q_2}...(\xi^m)^{q_m},
$$
with $a_{q_1...q_n}$ constant coefficients.
For such polynomials the following inequality takes place 
$$
|P(\xi^a)|\leq\sum_{q_1,..,q_m=0}^{N_1,..,N_m} |a_{q_1..q_m}||\xi^1|^{q_1} |\xi^2|^{q_2}...|\xi^m|^{q_m}.
$$
On the other hand, as $|\xi^a|\leq |\xi|=\sqrt{(\xi^1)^2+...+(\xi^m)^2}$, it follows that
$$
|P(\xi^a)|\leq Q(|\xi|)=\sum_{q_1,..,q_m=0}^{N_1,..,N_m} |a_{q_1..q_m}||\xi|^{q_1+q_2+..+q_m}.
$$
The right hand is a polynomial in the variable $|\xi|$ and it has positive coefficients. It is concluded that, for a given polynomial $P(\xi^a)$, there exists a polynomial $Q(\xi)$ in one variable such that
\be\lb{decime}
|P(\xi^a)|\leq Q(|\xi|).
\ee
In addition, the coefficients of $Q(\xi)$  are all positive, thus $Q(|\xi|)\geq 0$ for all values of $\xi^a$. The last property follows since the roots of $Q(\xi)$ are all negative, thus $Q(|\xi|)$ is positive and increasing. $\square$
\\

\emph{Proof of Proposition \ref{xcomp}:} The non linearity (\ref{descoteno})-(\ref{descoten3}) is not a polynomial expression in terms of the unknowns $u_\alpha=(g_{\mu\nu}, R, r_\gamma, c_{\mu\nu\alpha}, \widetilde{R}_{\mu\nu})$, therefore the previous lemma does not apply directly.  The problem resides in the factors in $f_q(u_\mu, \partial_\nu u_\mu)$ which contain the inverse metric $g^{\alpha\beta}$ and some of its derivatives. These expressions involve the inverse of the determinant
of the metric $|g_{\mu\nu}|$. This determinant is a polynomial $|g_{\mu\nu}|=P(g_{\alpha\beta})$ in the metric components $g_{\alpha\beta}$. In fact, it is not difficult to check that the inverse metric $g^{\alpha\beta}$ can be expressed in terms of $g_{\mu\nu}$ in the form
$$
g^{\alpha\beta}=\frac{P_{\alpha\beta}(g_{\mu\nu})}{|g_{\mu\nu}|},
$$
with $P_{\alpha\beta}$ a polynomial expression in its argument. This is a quotient of two polynomials which, in general, is not a polynomial. There are also Christoffel symbols $\Gamma^\alpha_{\beta\gamma}$ in the non linearities. The definition of the Christoffel symbol is
$$
\Gamma^\alpha_{\beta\gamma}=\frac{1}{2}g^{\alpha\epsilon}(\partial_\beta g_{\gamma\epsilon}+\partial_\gamma g_{\beta\epsilon}-\partial_\epsilon g_{\beta\gamma}).
$$
This expression also involves the inverse metric and  is of the form
$$
\Gamma^\alpha_{\beta\gamma}=\frac{P^\alpha_{\beta\gamma}(g_{\mu\nu}, \partial_\mu g_{\nu\alpha})}{|g_{\mu\nu}|},
$$
with $P^\alpha_{\beta\gamma}(g_{\mu\nu}, \partial_\mu g_{\nu\alpha})$ also a polynomial in its arguments.
In addition, there are also derivatives of the inverse metric $\partial_\gamma g^{\alpha\beta}$ in the non linearities in (\ref{descoteno})-(\ref{descoten3}), which are of the form
$$
\partial_\gamma g^{\alpha\beta}=\frac{P_{\alpha\beta\gamma}(g_{\mu\nu})}{|g_{\mu\nu}|^2}.
$$
These are essentially all the factors that contain the determinant $|g_{\mu\nu}|$ in the denominators.
It is obvious that $|g_{\mu\nu}|\to 0$ when $g_{\mu\nu}\to 0$. This implies that $f_q(x,t,0,0)\neq 0$ and in fact, this quantity may be divergent. However, there are finite negative powers $|g_{\mu\nu}|^{-m}$ in  $f_q(x,t, u_i, \partial_\mu u_j)$. Denote the maximal value of $m$ as $n$. Then, by  multiplying the system (\ref{descoteno})-(\ref{descoten3})  by $|g_{\mu\nu}|^n$ and by making a field redefinition $g^{\mu\nu}\to \overline{g}^{\mu\nu}=g^{\mu\nu}|g_{\mu\nu}|^n$ in the left hand side of (\ref{chic}),  the resulting system takes the form
\be\lb{chick2}
\overline{g}^{\mu\nu}\partial_{\mu}\partial_\nu u_q=\overline{f}_q(u_\mu, \partial_\nu u_\mu).
\ee
The resulting non linearity $\overline{f}_q(x,t, u_i, \partial_\mu u_j)=|g_{\mu\nu}|^ nf_q(x,t, u_i, \partial_\mu u_j)$ is a polynomial in the variables ($u_i$, $\partial_\mu u_j$), since the multiplication by $|g_{\mu\nu}|^n$ cancels out the negative powers $|g_{\mu\nu}|^{-m}$ that were present in $f_q(x,t, u_i, \partial_\mu u_j)$.  Furthermore, the resulting polynomial is such that $\overline{f}_q(0,0,x,t)=0$, since the non linearities  do not depend explicitly on $(x,t)$ and do not contain a constant term. All the derivatives of a polynomials are itself polynomials, and  satisfy the inequalities (\ref{decime}) with $Q_k(|u|)$ continuous and increasing, with $k$ the order of the derivative. Thus, the non linearity  $\overline{f}_q(u_\mu, \partial_\nu u_\mu)$ of the system (\ref{chick2}) is of x-compact support. $\square$
\\

The proposition given about is encouraging. However, despite the fact that there is a transformation of the system (\ref{descoteno})-(\ref{descoten3}) which converts it into (\ref{chick2}) with a non linearity $\overline{f}_q(u_\mu, \partial_\nu u_\mu)$ of x-compact support, there is no warrant that, once a solution is obtained, the global restrictions (\ref{cinf})-(\ref{cinf2}) for the modified inverse metric $\overline{g}^{\mu\nu}$ will be fulfilled. Thus, proposition \ref{stelle} may not apply due to this inconsistency.  Nevertheless, as shown below, this problem can be addressed by finding local solutions, valid in patches of the space-time manifold, and gluing them to a global one. 

\section{Conclusions}

The fact that the non linearity of the system (\ref{descoteno})-(\ref{descoten3}) is of x-compact support allows to make several conclusions about its solutions. The techniques to be employed below were applied in the book \cite{ringstrom} for the case of GR coupled to a real scalar field  $\varphi$. This is not the same situation as in the Stelle's gravity model. However, the system describing GR coupled to a scalar field and the system (\ref{descoteno})-(\ref{descoten3}) are both of the form  (\ref{chic}). For this reason, the results presented below are obtained partially by analogy with the approach of that book. Since there are several very technical details in \cite{ringstrom}, the following proofs will be just outlined, but the steps which are analogous and the ones which need to be modified will be indicated separately.

A first conclusion is that, given suitable initial conditions, there exists a $C^{\infty}$ solution for the Stelle's gravity model.

\begin{prop}\label{goril} Consider Stelle equations (\ref{gorila4})  in harmonic coordinates. There exists a global hyperbolic development  for the quantities $g_{(0)\mu\nu}=h_{(0)\mu\nu}+n_{(0)\mu}\otimes n_{(0)\nu}$, $k_{(0)\mu\nu}$, $G_{(0)\mu\nu}$ and $K_{(0)\mu\nu}$ defined in (\ref{initial}) if they are $C^\infty$ and satisfy the initial constraints (\ref{gorilap4}) and (\ref{gorilapp4}). The resulting solution is also $C^{\infty}$.
\end{prop}

\emph{Comment about the proof:} The proof of Proposition \ref{goril} is not a direct consequence of Proposition \ref{stelle}. The problem is that  Proposition \ref{stelle} ensures the existence of solutions $g_{\mu\nu}$ of of x-compact support. However, an space-time metric can not be of x-compact support. 
The source of this conflict are the global restrictions for the metric in (\ref{cinf})-(\ref{cinf2}), which may not be satisfied for a given solution $g_{\mu\nu}$. The only possibility is to apply this proposition in local patches $W_\alpha$ of the space-time manifold $M$.  These patches are selected such that the restrictions (\ref{cinf})-(\ref{cinf2}) are satisfied in $W_\alpha$. After obtaining these local solutions, an appropriate gluing procedure has to be implemented. This method is explained in detail in  \cite[Ch. 14]{ringstrom} and is applied for GR coupled to a scalar field. It can be generalized to the present situation, since the equations of motion of GR coupled to a scalar field are also of the form (\ref{chic}). The strategy is replace the quantity $\overline{g}^{\mu\nu}$ in (\ref{chick2}) by some quantities $A^{\mu\nu}$ with some suitable properties. The system them becomes
$$
A^{\mu\nu}(x,t,u)\partial_\mu\partial_\nu u_q(x,t)=f_q(x,t,u_i, \partial_\mu u_j),
$$
For instance, the quantities  $A_{\mu\nu}$ can be selected to be $A_{00}=\overline{g}_{00}$ in patches where  $\overline{g}_{\mu\nu}$ takes values in the interval
$[-3/2,-1/2]$, and have the property that the range of $A_{00}$ is contained in $[-2,-1/4]$. Analogous properties may be declared for $A_{0i}$ and $A_{ij}$.  There is nothing special about this choice of intervals, and a continuum of other choices are possible. The important point is however that the interval on which $A_{0i}=\overline{g}_{0i}$ should contain $0$. Moreover, the range of $A_{0i}$ should contain the interval on which $A_{0i}=\overline{g}_{0i}$, with a margin. 
A similar procedure is done with the other fields of the model, in this case $R$, $\widetilde{R}_{\mu\nu}$, $c_{\mu\nu\alpha}$, $r_\alpha$, together with a suitable modification of the initial and the harmonic coordinate conditions. After making these replacements, a solution is obtained. Given a point $p$, there is a neighbourhood $W_p$ of $p$ in which the solution obtained $g_{\mu\nu}$ coincides with the real space-time metric. One then consider the union of all these neighbourhoods
$\cup_{p}W_{p}=M$. There are further properties for the solution to be globally defined, which should be satisfied in the intersections $W_i\cap W_j$. Following  \cite[Ch. 14]{ringstrom} the following properties may be proved.
\\

- Given two solutions  $u_1$ and $u_2$ of the system (\ref{chick2}) corresponding to the neighbourhoods $W_1$ and $W_2$, then these are solutions of the Stelle's equations (\ref{gorila4}) in the intersection $W_1\cap W_2$.
\\

- The initial data induced by both solutions on $\Sigma\cap (W_1\cap W_2)$ coincide.
\\

- The solutions coincide in the intersection $W_1\cap W_2$.
\\

We have checked that the gluing procedure described in  \cite[Ch. 14]{ringstrom} can be applied to Stelle's quadratic gravity, when the equations of the model are formulated as a second order system of the form (\ref{chick2}). This is valid when harmonic coordinates are chosen. The detailed form of the arguments we employed are illustrated in the appendix of our reference \cite{nosotros}. These properties imply that the local solutions defined in $W_p$ may be glued to a global solution, which furthermore is smooth. The Proposition \ref{goril} then follows. $\square$
\\

The following assertion states that two different developments inducing the same initial data on $\Sigma$ arise as an extension of a common global hyperbolic development.

\begin{prop} Consider a given $C^\infty$  data $g_{(0)\mu\nu}=h_{(0)\mu\nu}+n_{(0)\mu}\otimes n_{(0)\nu}$, $k_{(0)\mu\nu}$, $G_{(0)\mu\nu}$ and $K_{(0)\mu\nu}$ for Stelle's gravity in an initial hypersurface $\Sigma$. Assume that there exists two hyperbolic developments ($M_a$, $g_a$)
and ($M_b$, $g_b$) inducing  this data with corresponding embeddings $i_a:  \Sigma\to M_a$ and $i_b:  \Sigma\to M_b$. Then there exists a global hyperbolic development
($M$, $g$) with a corresponding embedding $i:  \Sigma\to M$ and smooth orientation preserving maps $\psi_a: M\to M_a$ and $\psi_b: M\to M_b$, which are 
diffeomorphisms onto their images, such that $\psi_a^\ast g_a= g$ and $\psi_b^\ast g_b=g$. In addition 
$\psi_a\; \circ\: i=i_a$ and $\psi_b\;\circ\: i=i_b$. 
\end{prop}

\emph{Comments about the proof:} This proposition is similar to the one in \cite[Ch. 14]{ringstrom}. But, at first sight, its proof does not follows completely by analogy with the notions of that book.  The first apparent problem is the use of harmonic coordinates, which is a major technique employed in \cite{ringstrom}. In the Stelle's gravity models, the harmonic conditions are more restrictive that in standard GR coupled to a scalar field $\varphi$, due to the higher derivative nature of the former, as discussed in the section 3 given above. Furthermore, the description of harmonic coordinates the book \cite{ringstrom} employs is given in terms of a reference metric $h_{\mu\nu}$, which is also employed in the classic reference \cite{hawking}. 
For this reason, in the appendix A given below, the formulation of the Stelle's gravity model in harmonic coordinates, described in terms of a reference metric $h_{\mu\nu}$, is worked out explicitly. The resulting initial conditions are composed by the GR ones, together with new ones (\ref{menos}) and (\ref{uno})-(\ref{dos}). These last conditions are not required in the GR context, they are specific for the Stelle's quadratic gravity. The presence of the new constraints is an apparent complication for finding a proof of Proposition 7 by use of the procedures implemented in \cite{ringstrom}.

Nevertheless, these problems can be sorted out.  In order to see how, suppose that a solution ($M$, $g$) of the Stelle's equations of motion has been found. Assume  furthermore that this solution fulfils the harmonic condition in terms of a reference metric $h$, described in the appendix A. On the other hand, it is not assumed that the harmonic condition holds for ($M_a$, $g_a$) or ($M_b$, $g_b$).  At the moment, the only hyphotesis about ($M_a$, $g_a$) or  ($M_b$, $g_b$)  is that they are solutions of the Stelle's equations inducing the same initial data on the achronal hypersurface $\Sigma$ as ($M$, $g$). Thus, it is not assumed that these metrics are necessarily solutions of  (\ref{descoteno})-(\ref{descoten3}). Still, they are solutions of (\ref{gorila4}). The idea of the proof is then to construct a local diffeomorphism between $g$ and $g_a$ or $g_b$ and then to glue it to a global one.

Suppose now that a solution ($M$, $g$) has been constructed on an open set  $D\subseteq R\times \Sigma$. Assume furthermore that this solution satisfies the harmonic condition in terms of a reference metric $h_{\mu\nu}$ described in the appendix. Then, as shown in the formula (\ref{cove}) of the appendix  below, the relation $\Gamma_\mu=g_{\mu\nu}g^{\alpha\beta}\Sigma^\nu_{\alpha\beta}$ holds,  $\Sigma_{\alpha\beta}^\mu$ being the Christoffel symbols of the reference metric $h$ in a region $V\subseteq D$. This equality holds in any local coordinate system. This follows from the fact that the difference $$D_\mu=\Gamma_\mu-g_{\mu\nu}g^{\alpha\beta}\Sigma^\nu_{\alpha\beta},$$ is a 1-form and, if it is vanish in one coordinate system, it is identically zero.  Thus, the first task is to make a local choice of coordinates for ($M$, $g$). 

A convenient choice follows from the affirmation 12.5 of the book \cite{ringstrom}. Consider a point $p$ in the embedding $i_a(\Sigma)$. As $i_a(\Sigma)$ is a spatial surface in $M_a$, there exists coordinates $x^\mu$ in a region $U$ such that $x^\mu(p)=0$ and such that $q \in U\cap \Sigma$ if and only if $x^0(q)=0$. Furthermore $\partial_{x^0}|_q$ is a unit future director vector normal  to $\Sigma$ for $q\in U\cap \Sigma$. This is a local result and, more importantly, it is independent on the gravity model employed. Define $\hat{x}^i=x^i|_{U\cap i_a(\Sigma)}$, then $\hat{x}^i$ are coordinates on $U\cap i_a(\Sigma)$. Consider $\hat{y}^i=\hat{x}^i \circ\; i_a(\Sigma)$, then these are coordinates in $U_\Sigma=i_a^{-1}(U)$. Define $y^0=t$ and $y^i=\hat{y}^i$, then these are coordinates in $R\times U_{\Sigma}$.
In these terms,  one can make the replacement $\Gamma_\mu=g_{\mu\nu}g^{\alpha\beta}\Sigma^\nu_{\alpha\beta}$ in the equations of motion, when they are restricted to $V=R\times U_\Sigma\cap D$. Note that $g_{00}=-1$ and $g_{0i}=0$ for this choice of coordinates.

The next task is to construct a local coordinate system for $g_a$. Following the equations 14.20 to 14.22 of the book \cite{ringstrom} it follows  that there exists a local coordinate system $\widetilde{x}^\mu$ such that  \be\lb{katr}\Gamma^{(a)\mu}=\widetilde{g}_a^{\alpha\beta}\Theta^\nu_{\alpha\beta},\ee where $\Gamma_{\alpha\beta}^{(a)\mu}$ are the Christoffel symbols with respect to the metric $g_a$, and $\widetilde{g}_a^{\alpha\beta}$ are the inverse components metrics of $g^a_{\mu\nu}$, both referred to these $\widetilde{x}^\mu$ coordinates. In addition $\Theta^\nu_{\alpha\beta}$ are the Christoffel symbols of $h$ with respect to the $x$ coordinates. These coordinates are valid in a region $W$ specified in that reference. The deduction of this result does not include the new features 
(\ref{menos}) and (\ref{uno})-(\ref{dos}) outlined in the appendix and, therefore, is valid in the present context.

The discussion made so far is completely analogous to the one in  \cite[Ch. 14]{ringstrom}. However, care should be taken with the initial conditions, as the Stelle's equations are of fourth order while the Einstein equations are of second order. In order to clarify the differences that arise consider GR first. When dealing with the Einstein model, it is useful to define the patches $W_\Sigma=i_a^{-1}(W\cap \Sigma)$. The formulas of GR that are valid in these region are the following. For any point $q$ in this region one has that $y(q)=i_a\,\circ\,x(q)=i_a\,\circ\,\widetilde{x}(q)$. This formula leads to the following identity for the inclusions $$i_{a^\ast}\partial_{y^i}|_q=\partial_{\widetilde{x}^i}|_{i_a(q)}.$$ From this, it follows that $$g_{ij}(q)=\widetilde{g}_{aij}(i_a(q)),\qquad g_{\mu\nu}\,\circ\, y^{-1}=\widetilde{g}_{a\mu\nu}\,\circ\,x^{-1}.$$
In addition, for these coordinates \be\lb{blame}\widetilde{\Gamma}^{a}_\mu\,\circ\,x^{-1}=\Gamma_\mu\,\circ\,y^{-1},\ee and the second fundamental form satisfies the relation \be\lb{primaface}
k_{ij}(q)=\widetilde{k}_{aij}(i_a(q)).
\ee
This implies that $\partial_t g_{ij}(q)=\partial_{\widetilde{x}^0}\widetilde{g}_{aij}(i_a(q))$. As a conclusion it follows that \be\lb{primaface2}g_{\mu\nu}\,\circ\, y^{-1}=\widetilde{g}_{a\mu\nu} \, \circ\, x^{-1},\qquad k_{\mu\nu}\,\circ\, y^{-1}=\widetilde{k}_{a\mu\nu} \, \circ\, x^{-1}.
\ee
In addition $(\partial_{y^k}g_{\mu\nu})\,\circ\,y^{-1}=(\partial_{\widetilde{x}^k}\widetilde{g}_{a\mu\nu})\,\circ\,x^{-1}$, from where it obtained that 
 \be\lb{primaface3}(\partial_{t}g_{\mu\nu})\,\circ\,y^{-1}=(\partial_{\widetilde{x}^0}\widetilde{g}_{a\mu\nu})\,\circ\,x^{-1}.\ee
 The relations described above were employed in the GR relativity context. They show that both metrics $g$ and $g_a$ satisfy the same equations computed with their respective coordinates, and by use of uniqueness results for second order systems, it follows that the metric $g_a$ considered as a function of  $x$ has to coincide with $g$ considered as a function of $y$, in the region $W$.

 For the Stelle model instead, the use of harmonic coordinates requires the implementation of the new conditions (\ref{menos}) and (\ref{uno})-(\ref{dos}). These conditions involve the time derivatives $\dot{k}_{ij}$ and $\ddot{k}_{ij}$ of the second fundamental form $k_{ij}$. Thus, one is forced to derive identities such as (\ref{primaface})-(\ref{primaface3}) for $\dot{k}_{ij}$ or $\ddot{k}_{ij}$.  After these identities are obtained, one may conclude that with respect to the coordinates $\widetilde{x}$ the metric $g_a$ will satisfy the same equations that $g$ with respect to the $y^\mu$ coordinates. In addition, it may be argued that the initial data coincide when computed with their respective coordinates. 
However  the Stelle's equations (\ref{gorila4}) are of fourth order, while the arguments the book \cite{ringstrom} employs are related to second order quasi-linear hyperbolic systems. This implies that this line of reasoning do not apply directly to the present case.

The previous obstacle is of course discouraging. However, an approach to sort it out  comes from (\ref{blame}). This formula relates the quantity $\Gamma_\mu$ constructed in terms of $g$ with the quantity $\widetilde{\Gamma}^a_\mu$ constructed in terms of $\widetilde{g}_{a}$.  This fact, together with the independence of the harmonic description  in terms of the reference metric $h$ with respect of the coordinates ( cf. formula (\ref{katr})), imply that both $g$ and $\widetilde{g}_{a}$ satisfy the harmonic condition in the neighbourhood $W$. Therefore  both metrics $g$ and $\widetilde{g}_{a}$ are described by the second order quasi-linear system (\ref{descoteno})-(\ref{descoten3}) in $W$
with the remaining quantities $u_\alpha$ in (\ref{viktor}) computed with respect to their respective coordinates. The uniqueness arguments given in \cite{ringstrom} then apply for this system, since it is of second order. From this reasoning, it can be deduced after some work that, given a point $p$ in $W$, there exists an isometry $\psi_a$ in an open neighbourhood of $i^{-1}_a(p)$ with the property that $\psi_a^\ast g_a= g$. After further lengthy work following the steps of that book it can be shown that these local isometry can be glued to a global one, and the proposition will follow. $\square$
\\

The global hyperbolic development of the previous proposition may not be unique. Thus, it is of fundamental importance the notion of a maximal hyperbolic development. A  hyperbolic development ($M$, $g$, $\varphi$) is called maximal if, for any other global hyperbolic development ($M'$, $g'$, $\varphi'$),
there is an embedding $i':  \Sigma\to M'$ and an smooth orientation preserving maps $\psi: M'\to M$ 
such that $\psi^\ast g=g'$, $\psi^\ast \varphi=\varphi'$ and
$\psi\; o\: i'=i$.  
\\

\begin{prop} Given a valid initial data $g_{(0)\mu\nu}=h_{(0)\mu\nu}+n_{(0)\mu}\otimes n_{(0)\nu}$, $k_{(0)\mu\nu}$, $G_{(0)\mu\nu}$ and $K_{(0)\mu\nu}$ for the Stelle's equation, there exist a maximal global hyperbolic development, which is unique up to an isometry.
\end{prop}

\emph{Comment about the proposition:} The proof of this result is absolutely non trivial, but the techniques implemented in \cite[Ch. 16]{ringstrom} are valid in the present situation, since they involve abstract mathematical notions such as partially ordered sets, Zorn lemma or topology arguments. These arguments are not very sensitive to the details of the proof of the previous propositions, except for their statements. Based on this, the proposition then follows.
$\square$
\\

Finally, by analogy with \cite[Ch. 15]{ringstrom} the following conjecture may be formulated.
\\

\begin{prop}(Conjecture) Let ($M=\Sigma\times I$, $g$) a background solution of the vacuum Stelle gravity model. By denoting by  ($g_{(0)\mu\nu}$, $k_{(0)\mu\nu}$, $G_{(0)\mu\nu}$, $K_{(0)\mu\nu}$) the data induced
on $\{0\}\times \Sigma$ by the full solution, consider a sequence ($g_{(0)j\mu\nu}$, $k_{(0)j\mu\nu}$, $G_{(0)j\mu\nu}$, $K_{(0)j\mu\nu}$) of initial conditions converging to ($g_{(0)\mu\nu}$, $k_{(0)\mu\nu}$, $G_{(0)\mu\nu}$, $K_{(0)\mu\nu}$) for a suitable Sobolev norm, and satisfying the corresponding constraint equations. Then there exist $t_{1j}$ and $t_{2j}$ such that on $M_j=\Sigma\times (t_{1j}, t_{2j})$ 
there exists a Lorentzian metric $g_j$ which satisfy the Stelle's equation (\ref{gorila4}), and such that the initial data is 
 ($g_{(0)j\mu\nu}$, $k_{(0)j\mu\nu}$, $G_{(0)j\mu\nu}$, $K_{(0)j\mu\nu}$). The surface $\tau\times \Sigma$ is a Cauchy one when $\tau \in (t_{1j}, t_{2j})$. Furthermore, when $\tau\in I$, 
 the  data on such Cauchy hyper surface induced by $h_j$ converges to the one induced by $g$ for large $j$.
\end{prop}

The conjecture stated above is plausible sounding, but it may be not easy to prove it, since it appears that its proof is sensitive to the details of the theory. In GR coupled to a real scalar field $\varphi$, the suitable Sobolev space is $H^{l+1}$, with 
$2l>n+2$ being $n+1$ the space-time dimension. But in the present context, a suitable norm has to be found independently. Hopefully, the system (\ref{descoteno})-(\ref{descoten3}) obtained here may be helpful for these purposes. It would be a relevant task to come out with a proof (or a counterexample) of this assertion. In addition, it may be relevant to study the Cauchy problem when the hypothesis of global hyperbolicity is relaxed \cite{arefeva1}-\cite{arefeva2}. We leave this for a future investigation. 

\section*{Acknowledgments}
Both authors are supported by CONICET, Argentina. O.P.S is supported by the Beca Externa J\'ovenes Investigadores of CONICET. We gratefully acknowledge discussions with I. Ya. Arefeva and I. V. Volovich.  O.P.S warmly acknowledge the Steklov Mathematical Institute of the Russian Academy of Sciences in Moscow, where part of this work has been done, for their hospitality.

\appendix

\section{The description of harmonic coordinates in terms of a reference metric}

The use of a reference metric $h_{\mu\nu}$ in order to characterize harmonic coordinates, which was mentioned throughout  the text, can be described as follows  \cite{hawking}, \cite{ringstrom}. The expression for the Ricci tensor
corresponding to a generic metric $g_{\mu\nu}$ in an arbitrary coordinate system is given by 
\begin{equation}\label{descoh2}
R_{\mu\nu}=-\frac{1}{2}g^{\alpha\beta}\partial_{\alpha}\partial_{\beta}g_{\mu\nu}+P_{\mu\nu}(g,\partial g)+\nabla_{(\mu} \Gamma_{\nu)}.
\end{equation} 
The quantity $P_{\mu\nu}(g,\partial g)$ in (\ref{descoh2}) is not exactly equal to $Q_{\mu\nu}(g,\partial g)$ in (\ref{desco2}). However both expressions for the Ricci curvature are equivalent. Define the modified Ricci tensor  
\begin{equation}\label{descoh3}
\hat{R}_{\mu\nu}=-\frac{1}{2}g^{\alpha\beta}\partial_{\alpha}\partial_{\beta}g_{\mu\nu}+P_{\mu\nu}(g,\partial g)+\nabla_{(\mu} L_{\nu)}.
\end{equation} 
The quantities $L_\mu$ at the moment are not specified. But from the last two formulas it follows that
$$
\hat{R}_{\mu\nu}=R_{\mu\nu}+\nabla_{(\mu} D_{\nu)},
$$
where $D_\mu=L_\mu-\Gamma_\mu$. If the quantities $L_\mu$ are defined as
$$
L_\mu=g_{\mu\nu}g^{\alpha\beta}\Sigma^\nu_{\alpha\beta},
$$
with $\Sigma^\nu_{\alpha\beta}$ the Christoffel symbols of a reference metric $h_{\mu\nu}$\footnote{The reference metric $h_{\mu\nu}$ is not necessarily equal to the physical metric $g_{\mu\nu}$.}, then the difference $D_\mu=L_\mu-\Gamma_\mu$ is a 1-form. This property is of fundamental importance, since once $D_\mu=0$ in one coordinate system, then it will hold in any coordinates. In other words,  the equality
\be\lb{cove}
\Gamma_\mu=g_{\mu\nu}g^{\alpha\beta}\Sigma^\nu_{\alpha\beta},
\ee
will be valid in any local coordinate system.
Thus, given the initial surface $\Sigma$, if there is a domain $\Omega(\Sigma)$ in which $D_{\mu}$, $\nabla_\mu D_{\nu}$, $\nabla_\mu \nabla_\nu D_\alpha$ vanish, then they will vanish in the development $D(\Omega)$ described by the Stelle's equations (\ref{gorila4}). This fact will be independent on the choice of coordinates.

 The advantage of using a reference metric $h_{\mu\nu}$ is that the quantity $\Gamma_\mu$  is replaced by $L_\mu$, and the last expression involves second derivatives of the reference metric $h_{\mu\nu}$, and not the physical one $g_{\mu\nu}$. Thus, these terms do not spoil the quasi-linearity of the modified Ricci tensor $\hat{R}_{\mu\nu}$ and, as a consequence, $\hat{R}_{\mu\nu}$ becomes a quasi-linear second order expression for the metric $g_{\mu\nu}$. This is an important property employed during the text.

The reference metric $h_{\mu\nu}$ has not yet specified, and one has the freedom to make any choice. Usually, it is assumed that it has the Gaussian (synchronous) form
\be\lb{refer}
h=-dt^2+h_{ij}dx^i dx^j.
\ee
In what concerns the physical metric $g_{\mu\nu}$, one may fix the following initial conditions
\be\lb{ino1}
g_{ij}|_{t=0}=h_{ij}|_{t=0},\qquad g_{0i}|_{t=0}=0,\qquad g_{00}|_{t=0}=1.
\ee
\be\lb{ino2}
\partial_0 g_{ij}|_{t=0}=k_{ij}|_{t=0}.
\ee
In these terms, it follows that  
$$
D_0=L_0-\Gamma_0=L_0+\frac{1}{2}\partial_0 g_{00}+\text{Tr} \,K.
$$
where $K_{ij}=\partial_0 g_{ij}$ and, from the definitions above
$$
K_{ij}|_{t=0}=k_{ij}|_{t=0},$$   
$$
D_i=L_i-\Gamma_i=L_i+\frac{1}{2}\partial_0 g_{0i}+\frac{1}{2}g^{kl}(\partial_i g_{kl}-2\partial_k g_{il}) .
$$ 
The initial conditions are then $D_\alpha|_{t=0}=0$, $\partial_0D_\alpha|_{t=0}=0$ and $\partial^2_0D_\alpha|_{t=0}=0$, as the spatial derivatives are clearly zero initially. From (\ref{ino2}) it appears natural to impose the new constraints
\be\lb{menos}
\partial^2_0 g_{ij}|_{t=0}=\partial_0 k_{ij}|_{t=0},\qquad \partial^3_0 g_{ij}|_{t=0}=\partial^2_0 k_{ij}|_{t=0}.
 \ee
The last conditions are not employed in GR \cite{ringstrom} but they are natural identifications and may be required in the Stelle's model due to the higher order nature of the equations of motion (\ref{gorila4}).
Now, with all these assumptions at hand, the constraints $D_\alpha=0$ at the initial surface $\Sigma$ imply that
\be\lb{sero}
 \frac{1}{2}\partial_0 g_{00}|_{t=0}=-L_0|_{t=0}-\text{Tr} \,k|_{t=0},
  \ee
\be\lb{cero}
\frac{1}{2}\partial_0 g_{0i}|_{t=0}-=-L_i|_{t=0}-\frac{1}{2}g^{kl}(2\partial_k g_{il}-\partial_i g_{kl})|_{t=0}. 
\ee
These equations are standard in GR. Note that (\ref{ino1})-(\ref{ino2}) and (\ref{sero})-(\ref{cero}) specify the initial metric $g_{\mu\nu}$ values and their first time derivatives $\partial_t g_{\mu\nu}$. For a second order theory such as GR, these are enough. But for the Stelle model, the new constraints described below are required.
The constraints related to the first time derivatives of $D_\alpha$ are
$$
 \frac{1}{2}\partial^2_0 g_{00}|_{t=0}=-\partial_0 L_0|_{t=0}-\partial_0 \text{Tr} \,k|_{t=0}.
 $$
 \be\lb{uno}
\frac{1}{2}\partial^2_0 g_{0i}|_{t=0}-=-\partial_0 L_i|_{t=0}-\frac{1}{2}g^{kl}(2\partial_k k_{il}-\partial_i k_{kl})|_{t=0} -\frac{1}{2}k^{kl}(2\partial_k g_{il}-\partial_i g_{kl})||_{t=0}
\ee
This fixes the second time derivatives of the metric. The conditions related to the second derivatives of $D_\alpha$ are instead
$$
 \frac{1}{2}\partial^3_0 g_{00}|_{t=0}=-\partial^2_0 L_0|_{t=0}-\partial^2_0 \text{Tr} \,k|_{t=0},
 $$
\be\lb{dos}
\frac{1}{2}\partial^3_0 g_{0i}|_{t=0}-=-\partial^2_0 L_i|_{t=0}-\frac{1}{2}g^{kl}(2\partial_k \partial_0 k_{il}-\partial_i \partial_0 k_{kl})|_{t=0} -\frac{1}{2}k^{kl}(2\partial_k k_{il}-\partial_i k_{kl})||_{t=0}
\ee
$$
-\frac{1}{2}k^{kl}(2\partial_k  k_{il}-\partial_i  k_{kl})|_{t=0} -\frac{1}{2}\partial_0 k^{kl}(2\partial_k g_{il}-\partial_i g_{kl})||_{t=0}.
$$
The conditions (\ref{menos}) and (\ref{uno})-(\ref{dos}) are new features arising in the Stelle's gravity model, and fix the initial values of the metric up to third time derivatives. These are the conditions employed in the conclusions in section 6. Note that the description given here do not affect the validity of the harmonic coordinates as described in section 3, since the resulting equations are similar to
(\ref{id12})-(\ref{block2}) with $D_\mu$ playing a role analogous to $F_\mu$.


\begin{thebibliography}{plain}


 \bibitem{stelle}  K. S. Stelle, Phys. Rev. D 16
(1977) 953.
\bibitem{stelle2} K. S. Stelle, Gen. Rel. Grav. 9 (1978) 353.

\bibitem{noakes} D. Noakes  J. Math. Phys. 24 (1983) 7.

\bibitem{ringstrom}
H. Ringstrom \textit{The Cauchy Problem in General Relativity} European Mathematical Society, 2000.



\bibitem{ostrogradski} M.V. Ostrogradski, Mem. Acad. Imper. Sci. St. Petersbg., 6, 385 (1850).
\bibitem{pais} A. Pais and G.E. Uhlenbeck, Phys. Rev. 79, 145 (1950).

\bibitem{renor1} E. S. Fradkin and A. A. Tseytlin, Phys. Lett. 104 B, 377 (1981). 

\bibitem{renor2} E. S. Fradkin and A. A. Tseytlin,  Nucl. Phys. B 201 (1982) 469.


\bibitem{renor3} N. H. Barth and S. M. Christensen, Phys. Rev. D 28, 1876 (1983). 

\bibitem{renor4} I. G. Avramidi and A. O. Barvinsky,  Phys. Lett. 159 B, 269 (1985). 
\bibitem{odintsov}  I.L. Buchbinder, S.D. Odintsov, I.L. Shapiro \emph{Effective action in quantum gravity}  
Published in Bristol, UK IOP (1992).


\bibitem{sala1} A. Salam and J. A. Strathdee, Phys. Rev. D 18, 4480 (1978).

\bibitem{sala2} J. Julve and M. Tonin, Nuovo Cim. B 46, 137 (1978). 

 \bibitem{sala3} E. T. Tomboulis, Phys. Rev. Lett. 52 (1984) 1173.
 
  \bibitem{sala4} I. Antoniadis and E. T. Tomboulis,  Phys. Rev. D 33 (1986) 2756. 
  
  \bibitem{odintsov2} S. Nojiri, S. Odintsov and V. Okonomou Phys. Lett. B775 (2017) 44
  
  \bibitem{chenlim2} T. j. Chen, M. Fasiello, E. A. Lim and A. J. Tolley,  JCAP 1302, 042 (2013).

\bibitem{chenlim} T. Chen and M. Lim JCAP 1405, 010 (2014).

 

\bibitem{akitakob} Y. Akita and T. Kobayashi Modern Physics Letters A 31, 11 (2016) 1650067.

\bibitem{hinterblicher} K. Hinterbichler and M. Saravani Phys. Rev. D 93 (2016) 065006.

\bibitem{reno1} A. V. Smilga,  Nucl. Phys. B 706, 598 (2005).

 \bibitem{reno2} A. V. Smilga,  Phys. Lett. B 632, 433 (2006). 

\bibitem{reno3} J. F. Donoghue Phys. Rev. D 96, 044007 (2017).



\bibitem{clasico1} H. Lu, A. Perkins, C.N. Pope and K.S. Stelle,
Phys. Rev. Lett. 114 (2015) 171601.
\bibitem{clasico2} H. Lu, A. Perkins, C.N. Pope and K.S. Stelle, Phys. Rev. D 92 (2015) 12, 124019.

\bibitem{preciso1} K. Goldstein and J.J. Mashiyane, Phys. Rev. D 97, 024015 (2018).
\bibitem{preciso2} A. Perkins, “Static spherically symmetric solutions in higher derivative gravity,”
https://spiral.imperial.ac.uk:8443/handle/10044/1/44072.

\bibitem{clasico3} B. Whitt, Phys.
Rev. D 32 (1985) 379.
\bibitem{clasico4} Y.S. Myung, 
Phys. Rev. D 88 (2013) 2, 024039.

\bibitem{clasico5} W. Nelson, Phys. Rev. D 82 (2010) 104026.

\bibitem{clasico6} Y.S. Myung,
Phys. Rev. D 88 (2013) no.8, 084006.
\bibitem{clasico7} H. Liu, H. Lu and M. Luo,  Int. J. Mod. Phys.
D 21, 1250020 (2012).
\bibitem{clasico8} Z.Y. Fan and H. Lu, Phys. Rev. D 91 (2015) no.6, 064009.
\bibitem{clasico9} H. Lu and C.N. Pope, Phys. Rev. Lett. 106,
181302 (2011).
\bibitem{clasico10} S. Deser, H. Liu, H. Lu, C.N. Pope, T.C. Sisman and B. Tekin,  Phys. Rev. D 83, 061502 (2011).

\bibitem{lichnerowicz} 
H. Lu, A. Perkins, C.N. Pope and K.S. Stelle  Phys.Rev. D96 (2017) no.4, 046006. 

\bibitem{salvio} A. Salvio Front.in Phys. 6 (2018) 77.







 
 
 \bibitem{RelatividadChoquet}
	Y. Choquet-Bruhat. \textit{General Relativity and the Einstein Equations} Oxford Mathematical monographs 2009.

\bibitem{leray} J. Leray \textit{Hyperbolic Differential Equations} Institute for Advanced Study 1955.
\bibitem{foures} Y. Choquet-Bruhat Acta Math. 88 (1952) 141.

 
 \bibitem{Wald} R. Wald \textit{General Relativity} Chicago University Press 1984.
 
 \bibitem{hawking} S. Hawking and \textit{The Large Scale Structure of the Space-Time} Cambridge Monographs on Mathematical Physics 1973.
 
 \bibitem{Ehrlich} J. Beem, P. Ehrlich and K. Easley \textit{Global Lorentzian Geometry} CRC press 1981.
 
 \bibitem{O Neill} B. O Neill \textit{Semi-Riemannian Geometry with Applications to General Relativity} Academic Press 1983.
 
 \bibitem{gourgu} E. Gourgoulhon "Basis of numerical relativity" arXiv:gr-qc/0703035. Lectures given at the General Relativity Trimester held in the Institut Henri Poincare (Paris, Sept.-Dec. 2006) and at the VII Mexican School on Gravitation and Mathematical Physics (Playa del Carmen, Mexico, 26. Nov. - 2 Dec. 2006). 
 
 \bibitem{nosotros} J. Osorio Morales and O. Santill\'an "The existence of smooth solutions in q-models" arXiv:1808.07756, to appear in GRG.
 
 \bibitem{arefeva1} I. Arefeva, T. Ishiwatari and I. Volovich Theor. Math. Phys. 157 (2008) 1646.
 
 \bibitem{arefeva2} J. Friedman, M.S. Morris, I.D. Novikov, F. Echeverria, G. Klinkhammer, K.S. Thorne
and U. Yurtsever Phys. Rev.
D 42 (1990) 1915.
\end{thebibliography}
\end{document}